\documentclass[12pt]{article} 
\usepackage{amsmath, amsthm, amssymb,  calrsfs, wasysym, verbatim,  bbm, color, geometry, graphics, mathtools, indentfirst, enumitem,xcolor, url, multirow, lscape, subfigure}

\usepackage[square,numbers]{natbib}

\DeclareMathAlphabet{\mathcal}{OMS}{cmsy}{m}{n}

\title{A Two-Stage Method for Extending Inferences from a Collection of Trials}
\author{Nicole Schnitzler$^1$ and Eloise Kaizar$^2$}
\date{\footnotesize{$^1$ Ohio Colleges of Medicine Government Resource Center, The Ohio State University \\
$^2$ Department of Statistics, The Ohio State University}}

\begin{document}
\maketitle

\noindent \textbf{Abstract:} When considering the effect a treatment will cause in a population of interest, we often look to evidence from randomized controlled trials. In settings where multiple trials on a treatment are available, we may wish to synthesize the trials' participant data to obtain causally interpretable estimates of the average treatment effect in a specific target population. Traditional meta-analytic approaches to synthesizing data from multiple studies estimate the average effect among the studies. The resulting estimate is often not causally interpretable in any population, much less a particular population of interest, due to heterogeneity in the effect of treatment across studies. Inspired by traditional two-stage meta-analytic methods, as well as methods for extending inferences from a single study, we propose a two-stage approach to extending inferences from a collection of randomized controlled trials that can be used to obtain causally interpretable estimates of treatment effects in a target population when there is between-study heterogeneity in conditional average treatment effects. We first introduce a collection of assumptions under which the target population's average treatment effect is identifiable when conditional average treatment effects are heterogeneous across studies. We then introduce an estimator that utilizes weighting in two stages, taking a weighted average of study-specific estimates of the treatment effect in the target population. The performance of our proposed approach is assessed through simulation studies and an application to a collection of trials studying an online therapy treatment for symptoms of pediatric traumatic brain injury.

\vspace{0.5in}


\newpage

\section{Introduction}

Policymakers, clinicians, and others are often interested in learning what effect a treatment will cause in populations they serve. To address these types of questions, it would perhaps be ideal to have access to an internally valid randomized controlled trial (RCT) whose participants represent our target population. Unfortunately, such a study will often not be available and even internally valid RCTs often do not naturally extend to specific target populations due to treatment effect heterogeneity across populations \cite{buchanan2018generalizing, cole2010generalizing, hartman2015sample, westreich2017transportability}. When a collection of studies on the treatment are available, jointly analyzing the entire collection is likely to provide the best available information about the effect that a treatment would have in our target population. Traditional approaches to individual participant data (IPD) meta-analysis are designed to obtain an estimate in the population of study participants. Therefore, resulting estimates are often not causally interpretable in any meaningful population \cite{thompson2002howshould}. We propose a generalization of IPD meta-analytic methods that allows us to extend inferences from existing studies to obtain casually-interpretable estimates of treatment effects in our target population.

Traditional two-stage approaches to IPD meta-analysis synthesize information from collections of studies on the same treatment to obtain a single effect estimate. In the first stage, researchers consider the studies separately, obtaining effect estimates from each. These estimates are then treated as aggregate data and combined in the second stage, often via a weighted average \cite{burke2017meta, debray2015get}. These two-stage approaches are generally framed as either a fixed-effects (FEs) approach, typically derived from an assumption that there is a common true effect size across all of the studies, or a random-effects (REs) approach, derived instead from an assumption that the true study effects are drawn from some distribution \cite{debray2015get,higgins2009re}. We also follow the two-stage approach, where we estimate causally interpretable effect sizes in the first stage and incorporate study-to-study variability using a RE approach in the second stage.

For the first stage, methods have been developed to extend inferences from a single study to a target population. When extending from a single study, the goal is to adjust the study data to emulate the distribution of target population characteristics using ideas developed to balance treatment and control groups in traditional causal inference for observational data. Adopting the framework proposed by Rubin \cite{rubin1974estimating}, we consider an observation's potential outcome under a treatment to be the outcome that would be observed if the observation received that treatment. A study observation will typically only be observed under only one treatment, so their potential outcomes under all other available treatments are missing. The goal of causal inference can be thought of as estimating features of the distribution of these unobserved potential outcomes. When making inferences for a target population that is distinct from the population of study participants, we would like to learn about the distribution of the potential outcomes for observations in the target population, which were never observed since these individuals were never enrolled in a study or assigned to a treatment. Whereas traditional causal inference addresses the study observations' missing potential outcomes, extension also addresses the missing potential outcomes in the target population. In this paper, we focus on methods for extending inferences to a target population that has no overlap with the populations of study participants. This type of extension is often referred to as transportation (as opposed to generalization), but there has yet to be broad agreement on terminology and others have proposed different definitions \cite{dahabreh2019extending,dahabreh2020extendingmed, westreich2017transportability}.

Recently, these same ideas have been incorporated into methods for extending inferences from a collection of studies without explicitly addressing study-to-study heterogeneity. As discussed in more detail in later sections, these methods fully or partially pool studies' IPD, in a sense treating the data as though it had come from one large study instead of a collection of smaller ones. These methods would be natural under an assumption that participation in a particular study is not associated with an individual's expected treatment effect conditional on observed covariates \cite{dahabreh2019extending, dahabreh2020toward, dahabreh2019efficient}. Under this assumption, an individual's expected treatment effect, conditional on their covariates, would not change if they move from the target population to a study or from one study to another. In settings where there are between-study differences in research group, environment, trial design, etc., methods based on this assumption may not capture key features of the true treatment effect distributions. Vo et al.~\cite{vo2019novel,vo2021assessing} explores such heterogeneity for case-mix-adjusted meta-analysis. Our proposed approach incorporates such heterogeneity in the extending inferences setting, thus providing high-quality inference for a broader range of collections of studies by allowing for between-study treatment effect heterogeneity. 

To motivate our work in extending inferences from a collection of RCTs, we consider a meta-analysis of a collection of clinical trials studying the impact of an online therapy treatment on symptoms of pediatric traumatic brain injury (TBI). Pediatric TBI is one of the leading causes of morbidity and mortality in children and is the leading cause of acquired disability in children \cite{thurman2016epidemiology}. Online family problem-solving therapy (OFPST) is a intervention for children with pediatric TBI and their families that addresses cognitive appraisals, problem solving, and family communication \cite{wade2008brief}. It has been linked to reduced externalizing behaviors in adolescents after they experience a TBI \cite{wade2018online}. 

In this motivating example, we extended inferences from a collection of RCTs that studied the effect of OFPST on symptoms in children who had been hospitalized for moderate-to-severe TBI. These studies were conducted by different research groups, at different times, utilized different enrollment sites, and employed different eligibility criteria. Because of these study differences, we anticipate study-level heterogeneity in expected treatment effects. 
We extend inferences from this collection of four trials to a target population of children aged 11 to 13 years who had experienced a moderate-to-severe TBI within the last three months, as characterized by a cohort study of children who had experienced a moderate-to-severe TBI \cite{wade2002prospective}.

The rest of this paper is organized as follows. We first introduce relevant notation and inferential targets in Section \ref{Sec: Notation}. Weighted methods for extending inferences that pool studies' IPD are reviewed in Section \ref{Sec: Existing}. Then, in Section \ref{Sec: TwoStage}, we introduce a two-stage approach that allows for between-study heterogeneity in treatment effects. We consider the performance of approaches to extending inferences from a collection of studies using simulation studies in Section \ref{Sec: Simulations}. Application to our motivating example is presented in Section \ref{Sec: TBI}. Finally, Section \ref{Sec: Discussion} summarizes our work, discusses its implications, and considers potential areas of future work. 

\section{Notation and Inferential Targets} \label{Sec: Notation}

We assume that we have IPD from a collection of $m$ RCTs, indexed by $s \in \mathcal{S} = \{1,2,3,...,m\}$. Each considers a pair of treatments $\mathcal{A} = \{a,a'\}$ where $a$ represents the treatment and $a'$ represents the control. In addition, we assume we have an SRS of the baseline covariates from an infinite target population of interest that does not include any study observations. We suppose the combined RCT and population data have $n$ observations indexed by $i=1,2,...,n$. Let $S_i \in \{0,\mathcal{S}\}$ be a random variable denoting the study membership for observation $i$ with $S_i=0$ denoting membership in the target population. Then, $n_s= \sum_{i=1}^n I(S_i=s)$ for $s \in \{0,\mathcal{S}\}$ is the sample size for each data source and $n = \sum_{s=0}^m n_s$. Throughout this paper, we use the notation $f(\cdot)$ to denote densities. 

For each RCT study participant $i : S_i \ne 0$, we observe variables including a study indicator, $S_i$, treatment assignment, $A_i$, an outcome of interest, $Y_i$, and a vector of baseline covariates, $\boldsymbol{X}_i$. Within each RCT, these vectors are mutually independent and identically distributed. Because observations in our target population are not enrolled in a study, for each observation in our sample from the target population we observe only the study indicator $S_i = 0$ and a vector of baseline covariates $\boldsymbol{X}_i$. 

To aid our focus on estimating causal effects, we adopt the notation of Rubin \cite{rubin1974estimating} and let $Y_i^a$ be the potential outcome for observation $i$ under treatment $a$. This is the outcome that would be observed for observation $i$ if they received treatment $a$. We focus on two common inferential targets when extending inferences: (1) the target population's average treatment effect (TATE), 
\begin{equation} \label{Eq: Def_TATE}
    \Delta = E\left(Y_i^a - Y_i^{a'}|S_i=0\right), 
\end{equation}
i.e., the expected treatment effect for a randomly selected individual in the target population, and (2) the mean potential outcome under $a \in \mathcal{A}$ in the target population, $\mu_a = E(Y_i^a|S_i=0)$, i.e., the expected outcome under treatment $a \in \mathcal{A}$ for a randomly selected individual in the target population. Notice that $\Delta = \mu_a - \mu_{a'}$, so estimators of $\Delta$ can often be viewed as a difference between estimates of mean potential outcomes. 

\section{Existing Methods for Extending Inferences} \label{Sec: Existing}

Recently, methods for extending inferences that can be thought of as fully or partially pooling studies' IPD have been proposed \cite{dahabreh2020toward, dahabreh2019efficient}. By effectively pooling the study data (after accounting for potentially different treatment assignment strategies), these methods, in a sense, treat the data as though they had come from one large study instead of a collection of smaller ones. Versions of these methods include approaches that rely on modeling the conditional average treatment effect (ATE) in the pooled study population, i.e, for $i: S_i \in \mathcal{S}$, (e.g., see \cite{dahabreh2020toward, dahabreh2019efficient}) and adaptations for settings where covariate data are systematically missing from some of the studies in $\mathcal{S}$ (e.g., see \cite{steingrimsson2022systematically}). Such methods could build on weighted, model-based, or doubly-robust/augmented methods for extending inferences from a single study. In this paper, we focus on weighted methods for estimating the TATE in an infinite target population.

\subsection{Assumptions} \label{SubSec: PooledAssump}

When extending inferences, we must make a collection of assumptions that enable us to obtain causally interpretable estimates from the study data and to extend those estimates to our target population. The specifics may vary by the inferential target and method of estimation, but generally involve a collection of one assumption from each of five categories. 

The first three categories consist of variations of usual the causal inference assumptions about the consistency and exchangeability of potential outcomes and the positivity of treatment assignment. Dahabreh et al. \cite{dahabreh2020toward, dahabreh2019efficient} propose the following assumptions for these categories:
\begin{itemize}[label={}]
    \item \textbf{A1}. If $A_i =a$, then $Y_i^a = Y_i$ for all $i=1,...,n$ and $a \in \mathcal{A}$. 
    \item \textbf{A2}. For all $s\in \mathcal{S}$, for all $a \in \mathcal{A}$, and for $\boldsymbol{x}$ such that $f(\boldsymbol{x},S=s) > 0$, $$E(Y^a| \boldsymbol{X}=\boldsymbol{x},S=s, A=a) = E(Y^a|\boldsymbol{X}=\boldsymbol{x},S=s).$$
    \item \textbf{A3}. For all $s\in \mathcal{S}$, for all $a \in \mathcal{A}$, and for $\boldsymbol{x}$ such that $f(\boldsymbol{x},S=s) > 0$, $$P(A=a|\boldsymbol{X}=\boldsymbol{x},S=s) >0.$$
\end{itemize}
Assumption A1 implies consistency of the potential outcomes regardless of study membership. Exchangeability of the potential outcomes is implied by Assumption A2, which states that, conditional on their baseline covariates, an individual's mean potential outcome in a given study does not depend on their treatment assignment. Finally, Assumption A3 states that within each study there are no sets of possible baseline covariates that would prevent an individual from receiving any of the treatments being considered. 

The fourth and fifth categories are exchangeability and positivity assumptions to enable extension from the studies to the target population. Dahabreh et al. \cite{dahabreh2020toward, dahabreh2019efficient} propose the following: 
\begin{itemize}[label={}]
    \item \textbf{A4}. For all $s \in \mathcal{S}$, for all $a, a' \in \mathcal{A}$ and for $\boldsymbol{x}$ such that $f(\boldsymbol{x},S=0) > 0$, $$E(Y^a - Y^{a'}|\boldsymbol{X}=\boldsymbol{x},S=0) = E(Y^a - Y^{a'}|\boldsymbol{X}=\boldsymbol{x},S=s).$$
    \item \textbf{A5}. For all $s\in \mathcal{S}$ and for $\boldsymbol{x}$ such that $f(\boldsymbol{x},S=0) > 0$, $$P(S=s|\boldsymbol{X}=\boldsymbol{x})>0.$$
\end{itemize}
Assumption A4 states that individuals' ATEs, conditional on their covariate values, are the same across all of the studies and the target population. Under this assumption, an individual's treatment effect is conditinally independent of their study membership. This implies that study membership does not need to be accounted for except to allow for differences in the treatment assignment mechanism across studies. Therefore, data from the RCTs can be pooled (after accounting for treatment assignment) when extending inferences if this assumption holds. Assumption A5 requires sufficient overlap between the covariate distribution in the target population and each individual study. When the study data are pooled, Dehabrah et al. note that we can relax this assumption and only require sufficient overlap between the target population and the pooled study data \cite{dahabreh2020toward, dahabreh2019efficient}. 

\subsection{Weighted Estimators}

One type of existing estimator of the TATE based on observations from the (partially) pooled study sample is a weighted average. These estimators take the form 
\begin{equation} \label{Eq: Pooled_WtEst}
\begin{split}
    \hat{\Delta}  = & \sum_{i=1}^n\left(\frac{\hat{w}_i(a)}{\sum_{i=1}^n \hat{w}_i(a)} - \frac{\hat{w}_i(a')}{\sum_{i=1}^n \hat{w}_i(a')}\right)Y_i \\
\end{split}
\end{equation}
where, for each $a \in \mathcal{A}$, $\hat{w}_i(a) \geq 0$ with $\hat{w}_i(a) = 0$ if $S_i=0$ or $A_i \ne a$. By carrying through the sum, it is easy to see that $\hat{\Delta}$ is the difference between weighted averages of the treated and control units in the study sample. These weighted averages could be used to estimate $\mu_a$ and $\mu_{a'}$. Because existing weighted methods (partially) pool the study data, the estimator of TATE in Equation \ref{Eq: Pooled_WtEst} has the same form as estimators proposed for extension from a single study (e.g., see \cite{buchanan2018generalizing,dahabreh2019extending}). Specific estimators are differentiated by different choices for the estimated weights, $\{\hat{w}_i(\cdot)\}_{i=1}^n$. 

A na\"{\i}ve approach would set the estimated weights to be a simple indicator of receiving treatment in any study
\begin{equation} \label{Eq: SingleStudy_SimpleWts}
    \hat{w}_{ua,i}(a) = I(A_i=a)I(S_i\ne0)
\end{equation}
for each $a \in \mathcal{A}$, where the $ua$ stands for unadjusted. The resulting estimator, $\hat{\Delta}_{ua}$  is the difference between the average outcomes among the treated units across all studies and the control units across all studies. These simple weights are appealing because of their computational ease, but $\hat{\Delta}_{ua}$ does not adjust for confounding due to treatment assignment or study membership. This lack of adjustment means that $\hat{\Delta}_{ua}$ will only be a consistent estimator of the TATE in special cases such as when the pooled study data is a simple random sample from the target population, potential outcomes are consistent and independent of both treatment assignment and study, and treatment assignment is independent of both baseline covariates and study membership. 

To adjust for the potential confounding due to both treatment assignment and study membership, Dahabreh et al.~\cite{dahabreh2020toward} propose the following estimated weights for each $a \in \mathcal{A}$, 
\begin{equation} \label{Eq: MultStudy_PooledWts}
    \hat{w}_{pool,i}(a) = \frac{I(A_i=a)}{\hat{e}_a(\boldsymbol{X}_i,S_i)} \frac{\hat{p}(\boldsymbol{X}_i,0)}{1-\hat{p}(\boldsymbol{X}_i,0)}I(S_i \ne 0)
\end{equation}
where $\hat{p}(\boldsymbol{X}_i,0)$ and $\hat{e}_a(\boldsymbol{X}_i,S_i)$ are estimates of $P(S=0|\boldsymbol{X}_i)$ and $P(A=a|\boldsymbol{X}_i,S_i)$, respectively,. These estimated weights utilize both inverse-probability of treatment weights (IPTWs) and inverse odds of study participation weights (IOSPWs). The IPTWs, $\frac{I(A_i=a)}{\hat{e}_a(\boldsymbol{X}_i,S_i)}$, weight each observation by the inverse if its propensity score, the probability of receiving the treatment it received, and are used to adjust for potential confounding due to treatment assignment. The IOSPWs, $\frac{\hat{p}(\boldsymbol{X}_i,0)}{1-\hat{p}(\boldsymbol{X}_i,0)}$, weight each observation by the inverse of its odds of being enrolled in any study, as opposed to the target population. These are used to adjust for confounding due to membership in any study. Under Assumptions A1 to A5, the resulting estimator, $\hat{\Delta}_{pool}$, is a consistent estimator of the TATE so long as $\hat{p}(\boldsymbol{X},0)$ and $\hat{e}_a(\boldsymbol{X},S)$, for each $a \in \mathcal{A}$, are consistent estimators of corresponding probabilities \cite{dahabreh2020toward, dahabreh2019efficient}. 

A more recently proposed method for causal interpretable meta-analysis does not require pooling study data \cite{clark2023causally}. This proposed method utilizes the same estimator as we propose in Section \ref{Subsec: TS_Est}, but considers the distinct estimand of the expected treatment effect we would see \emph{if we could run a new trial in the target population}. They do this by conceptualizing the conditions experienced by an individual receiving some treatment $a \in \mathcal{A}$ in study $s \in \mathcal{S}$ as a random variable $K_s^a(a)$ such that $K_1^a(a),...,K_m^a(a)$ are an i.i.d sample from some distribution. They then follow the lead of VanderWeele and Hernán \cite{vanderweele2013causal} to consider the two-dimensional treatment potential outcome for an observation in study $s \in \mathcal{S}$ under treatment $a \in \mathcal{A}$ as $Y(a,k_s^a)$. This approach contrasts from ours, where we consider the study akin to a covariate rather than a treatment variation. These distinctions lead to differences between associated assumptions and operating characteristics in our proposed method and the method proposed by Clark et al.~\cite{clark2023causally}. 

\section{A Two-Stage Approach to Extending Inferences} \label{Sec: TwoStage}

Existing methods for extending inferences that (partially) pool studies' IPD do not allow for between-study treatment effect heterogeneity. In this section, we propose a two-stage method for extending inferences from a collection of studies that allows for participation in a particular study to impact conditional ATEs. In Section \ref{Subsec: TS_Assumption}, we propose a collection of identifiability assumptions. We demonstrate that the TATE can be identified under these assumptions in \ref{Subsec: TS_ID}. Then, we introduce our proposed two-stage weighted approach for inferring TATEs in Section \ref{Subsec: TS_Est}. 

\subsection{Assumptions} \label{Subsec: TS_Assumption}

To extend inferences under our two-stage approach in settings with between-study heterogeneity, we must first choose a set of assumptions that represent the five categories introduced in Section \ref{SubSec: PooledAssump}. We directly adopt three of the assumptions proposed by Dahabreh et al. \cite{dahabreh2020toward, dahabreh2019efficient}: A2, A3, and A5. But, assumptions A1 and A4 restrict inferences to settings where individuals' potential outcomes and conditional ATEs are not associated with study. Thus, we propose alternative assumptions for these categories.

Because assumption A1 does not take into account whether an individual is enrolled in a study or is a member of the target population, it implies that an individual's potential outcome under a given treatment would not change if they were moved from one study to another. It is easy to imagine settings where this assumption might not be reasonable. For example, consider two RCTs of the same treatment that were conducted at hospital systems in two different communities. We might expect see better outcomes among some groups of patients in one hospital than the other because of differences in community and hospital-level characteristics. As an alternative, we assume
\begin{itemize}[label={}]
    \item \textbf{B1}. For all $s \in \mathcal{S}$, $a \in \mathcal{A}$, and $i \in \{1,2,...,n:S_i=s\}$, if $A_i=a$ then $Y_i^a = Y_i$. 
\end{itemize}
By conditioning consistency on study membership, we allow for settings where an individual’s potential outcome under their assigned treatment changes if their study membership changes. Note also that we do not specify any relationship between the potential and observed outcomes in the target population ($s=0$). Such specification is unnecessary because we do not observe outcomes for members of the target population.

Under Assumption A4, individuals' ATEs, conditional on their covariates would not change if they were moved from the target population to a study or from one study to another. This does not allow for any study-level treatment effect heterogeneity. We relax this assumption and instead assume that the conditional ATEs in each study follow some distribution with a mean at the conditional ATE in the target population. The resulting assumption is 
\begin{itemize}[label={}]
    \item \textbf{B4}. For all $a, a' \in \mathcal{A}$ and for $\boldsymbol{x}$ such that $f(\boldsymbol{x},S=0) > 0$, 
    \begin{equation*}
        E(Y^a - Y^{a'}|\boldsymbol{X}=\boldsymbol{x},S=0) = E_{S}\left[E(Y^a - Y^{a'}|\boldsymbol{X}=\boldsymbol{x},S)\big|S \in \mathcal{S} \right].
    \end{equation*}
\end{itemize}
This new assumption allows for between-study heterogeneity in the conditional treatment effects.

Our final collection of identifiability assumptions consists of assumptions B1, A2, A3, B4, and B5. Moving forward, we will call this set of assumptions ``Assumption $\mathcal{B}$" for short. 

\subsubsection{Identifiability} \label{Subsec: TS_ID}

To obtain causally interpretable estimates of the TATE, we must be able to write it in terms of features that can be estimated from observed data. When such an expression is possible, we say that the TATE is identifiable. Under assumption $\mathcal{B}$ and a regularity condition, the TATE, as defined in Equation \ref{Eq: Def_TATE}, is identifiable. 

\vspace{1ex}

\noindent \textbf{Theorem}: \emph{Under regularity condition $$E_S\left\{E_X\left[\left|E_{Y^a,Y^{a'}}\left(Y^a - Y^{a'}\big|\boldsymbol{X},S\right)\right|\big|S=0 \right]\big|S \in \mathcal{S} \right\} < \infty$$ and Assumption $\mathcal{B}$}, 
\begin{equation} \label{Eq: Identfiability1}
\begin{split}
    \Delta \equiv E(Y^a-Y^{a'}|S=0) = E_S\left(\Delta_S|S \in \mathcal{S}\right)
\end{split}    
\end{equation}
\emph{with} 
\begin{equation} \label{Eq: Identfiability2}
\begin{split}
     \Delta_s \equiv & E_X\left[E_{Y^a,Y^{a'}}(Y^a-Y^{a'}|\boldsymbol{X},S=s)\big|S=0\right] \\
     = & E_{A,X,Y}\left(\left[\frac{w(a,s)}{E_{A,X}\left(w(a,s)|S=s\right)} - \frac{w(a',s)}{E_{A,X}(w(a',s)|S=s)}\right]Y\Big|S=s\right)
\end{split}
\end{equation}
\emph{for each $s \in \mathcal{S}$ where
\begin{equation*}
    w(a,s) = \frac{I(A=a)}{P(A=a|\boldsymbol{X}, S=s)}\frac{P(S=0|\boldsymbol{X})}{P(S=s|\boldsymbol{X})}.
\end{equation*}
for each $s \in \mathcal{S}$ and $a \in \mathcal{A}$.} 

\vspace{1ex}

\noindent A detailed proof of this result can be found in Appendix A. 

\subsection{Estimator} \label{Subsec: TS_Est}

Based on the identifiability result in Equation \ref{Eq: Identfiability1}, we propose a class of estimators of the TATE that consist of weighted averages of study-specific estimates of the TATE, $\hat{\Delta}_s$, $s\in \mathcal{S}$. Specifically, we propose estimating the TATE with the two-stage estimator 
\begin{equation} \label{Eq: TwoStage_Est}
    \hat{\Delta}_{two} = \frac{1}{\sum_{s=1}^m w_s}\sum_{s=1}^m w_s\hat{\Delta}_s.
\end{equation}
where the $w_s$, $s \in \mathcal{S}$, are strictly positive study-level weights. This estimator is similar to the weighted estimator typically used in two-stage IPD meta-analyses but has one key difference. Unlike these more traditional meta-analytic approaches, $\hat{\Delta}_{two}$ utilizes causally interpretable study-specific estimates of the effect of a treatment in a specific target population. 

Based on the identifiability result in Equation \ref{Eq: Identfiability2}, we propose utilizing the study-specific weighted estimator of the TATE for each $s \in \mathcal{S}$
\begin{equation} \label{Eq: StudySpec_Est}
    \hat{\Delta}_s =  \sum_{i: S_i=s} \left(\frac{\hat{w}_i(a,s)}{\sum_{i: S_i=s} \hat{w}_i(a,s)}-\frac{\hat{w}_i(a',s)}{\sum_{i: S_i=s} \hat{w}_i(a',s)}\right)Y_i,
\end{equation}
where 
\begin{equation} \label{Eq: StudySpec_Est_wts}
    \hat{w}_i(a,s) = \frac{I(A_i=a)}{\hat{e}_a(\boldsymbol{X}_i,s)}\frac{\hat{p}(\boldsymbol{X}_i,0)}{\hat{p}(\boldsymbol{X}_i,s)},
\end{equation}
$e_a(\boldsymbol{x},s) = P(A=a|\boldsymbol{X}=\boldsymbol{x},S=s)$, and  $p(\boldsymbol{x}, s) = P(S=s|\boldsymbol{X}=\boldsymbol{x})$ for each $a \in \mathcal{A}$ and $s \in \mathcal{S}$ with hats indicating estimates of these quantities. This study-specific estimator has a similar form to the pooled weighted estimator, $\hat{\Delta}_{pool}$, weighting each observation in study $s$ by a study-specific IPTW and the inverse of its odds of being enrolled in the study as opposed to the target population. Each member of this class of estimators is distinguished by the specification of the study-level weights and the methods used to estimate the study-specific propensities for treatment assignment and study membership. 

We choose to simplify our presentation and theoretical derivations by using parametric binomial and multinomial logistic regression models to estimate the study-specific propensities. Similar approaches have been used in prior work developing methods for extending inferences (e.g., \cite{dahabreh2019extending}). In practice, one could choose some other parametric or nonparametric method. To allow for different treatment assignment mechanisms across studies, we propose fitting separate logistic regression models for treatment within each study to obtain estimates of each individual's propensity of treatment within their assigned study, $e_a(\boldsymbol{X}_i,S_i)$, noting that $e_{a'}(\boldsymbol{X}_i,S_i) = 1- e_a(\boldsymbol{X}_i,S_i)$. To simultaneously estimate the conditional probabilities of study membership, $p(\boldsymbol{x}, s)$, $s \in \{0,\mathcal{S}\}$, we propose fitting a single multinomial logistic regression model for study membership, where membership in the target population is considered to be the baseline category. We do not consider random-effects study-specific or study membership models because the primary purpose of the unit-level weights is to obtain balance \cite{schuler2016propensity}. 

To utilize our proposed two-stage approach, we must also choose a form for the study-level weights. These weights are used to combine the study-specific estimates and determine which studies have the most, or least, impact on the final estimate of the TATE. The statistical properties of the two-stage estimator depend on this choice of study-level weights. We conjecture that for carefully chosen study-level weighting schemes, $\hat{\Delta}_{two}$ can be a consistent estimator of the TATE. A detailed outline of our reasoning for the consistency of our proposed two-stage estimator for evenly weighted study-specific estimates under Assumption $\mathcal{B}$ and some regularity conditions can be found in Appendix B.

\subsubsection{Variance Estimation} \label{SubSec: Var_Est}

To obtain variance estimates and confidence intervals for $\hat{\Delta}_{two}$, we propose a stratified bootstrap procedure that mimics our imagined data generating procedure by resampling from the studies and the target population separately. We first take a bootstrap sample from the target population sample by simply sampling with replacement, maintaining the original sample size. Second, we obtain a bootstrap sample of the study data by following a two-stage procedure. To capture study-level heterogeneity in treatment effects, we first sample with replacement from $\mathcal{S}$ to get a bootstrap sample of $m$ studies, denoted $\{b_1,...,b_m\}$. We generate 
$n_j(a)$ treatment and $n_j(a')$ control individuals in each bootstrapped study $b_j$, $j=1,2,...,m$, where $n_j(a)$ and $n_j(a')$ are the respective sample sizes in the original study $j$. To do so, we sample units with replacement separately from each arm. In each combined bootstrap sample, we calcluate a TATE estimate and use the resulting sample variance to estimate the variance of $\hat{\Delta}_{two}$ \cite{chen2017variance}. We use percentile intervals, which are the $\frac{\alpha}{2}$ and $1-\frac{\alpha}{2}$ quantiles of the bootstrap estimates, to obtain approximate $100(1-\alpha)\%$ confidence intervals \cite{wasserman2006all}. 

\section{Simulations} \label{Sec: Simulations}

To examine the performance of our proposed two-stage estimator of the TATE, we conducted proof-of-concept simulations. 

\subsection{Data Generation}

We considered simulations with three studies, a realistic number given that we need each study to consider the same treatment and have available IPD, and thirty studies, to approximate a large sample setting. Simulations with three studies contained $n=10,000$ individuals while simulations with thirty studies contained $n=50,500$. In each setting considered, the portion sampled from the target population contained around 5,500 participants on average.

In all settings considered, we used the following general data generation process for each of 1,000 replications. 
\begin{enumerate}
    \item For $n$ observations, $i=1,2,...,n$, a covariate $X_i$ was drawn from a standard normal distribution. 
    \item Observations were assigned to either enrollment in a study $s \in \mathcal{S} = \{1,2,...,m\}$ or to membership in the target population using a multinomial logistic regression model where 
     \begin{equation} \label{Eq: Sim_SMGen}
    \begin{split}
        & P(S_i=s|X_i) = \text{expit}\left(\beta_{s,0} +X_i \beta_{s,1}\right) \text{, for } s =1,2,...,m \\
        & P(S_i=0|X_i) = 1 - \sum_{s=1}^{m}P(S_i=s|X_i).
    \end{split}
    \end{equation}
    (Recall that the expit function is the inverse of the logit function.) Values for $\beta_{s,1}$, $s \in \mathcal{S}$, were chosen in the range of approximately -0.4 to 0 so the covariate distributions differed across $s \in \{0,\mathcal{S}\}$, with the distribution of $X$ in the target population generally centered at or above the centers of the distributions in the studies and the centers of the study distributions increasing from Study 1 to Study $m$. Exact values are reported in Appendix C. Values for $\beta_{s,0}$, $s \in \mathcal{S}$, were chosen to achieve desired sample size patterns across the studies, as discussed below.
    \item Within each study, we implemented a completely randomized treatment assignment mechanism with $P(A_i=1|S_i=s) =0.5$ for all $s \in \mathcal{S}$.
    \item Outcomes were generated according to the outcome generation model
     \begin{equation} \label{Eq: Sim_OutGen}
    \begin{split}
        Y_i = &  \nu_{S_i} + A_i\gamma_{S_i} + X_i\lambda_{S_i} + A_iX_i\kappa_{S_i} + \epsilon_{i} \\
    \end{split}
    \end{equation}
     where $\epsilon_i \overset{iid}{\sim} N(0,1)$. For all $i=1,2,...,n$, $\epsilon_i$ was mutually independent of $X_i$, $A_i$, and $S_i$. Under this model, the conditional ATE for an observation in $s \in \{0,\mathcal{S}\}$ is 
    \begin{equation} \label{Eq: CondATE}
        E(Y^1-Y^0|S=s, X=x) = \gamma_s + \kappa_sx
    \end{equation}
    For all $s \in \{0,\mathcal{S}\}$, $\boldsymbol{\theta}_s = (\nu_s, \gamma_s, \lambda_s,\kappa_s)^t$ and we assumed $\boldsymbol{\theta}_0 = (-1,-1,0.5, -0.5)^t$ so that the true TATE was 
    \begin{equation} \label{Eq: TreuTATE}
        E(Y^1-Y^0|S=0) = -1 - 0.5E(X|S=0).
    \end{equation}
\end{enumerate}

\begin{table}[t]
    \centering
    \footnotesize
    \begin{tabular}{c|cc}
    \hline
       \multirow{2}{*}{Setting} & \multirow{2}{*}{Study Sizes} & Main Source of Treatment \\
        & & Effect Heterogeneity \\
        \hline
        \hline
        1 & Similar & Measured Features ($\Sigma_{\gamma} < \Sigma_{\kappa})$\\
        \hline
        2 & Different & Measured Features ($\Sigma_{\gamma} < \Sigma_{\kappa})$\\
        \hline
        3 & Different & Unmeasured Features ($\Sigma_{\gamma} > \Sigma_{\kappa})$\\
        \hline
    \end{tabular}
    \caption{General simulation settings considered in both three study and thirty study settings.}
    \label{Table: Simulation Settings}
\end{table}

We considered few settings for the distribution of study sizes and the outcome generation model, as summarized in Table \ref{Table: Simulation Settings}. For study sizes, we considered settings where studies had a similar number of participants (around 1,500 each) or different numbers, increasing from Study 1 (265 for m=3, 675 for m=30) to Study $m$ (2,475 for m=3, 2,988 for m=30). We utilized a numeric algorithm to solve for values of $\beta_{0,s}$, $s=1,2,..,m$, needed in the study membership generation model (Equation \ref{Eq: Sim_SMGen}) to obtain the desired study and target sample sizes, on average \cite{robertson2022using}. These target sizes are reported in Appendix C.

For treatment effect heterogeneity, we allowed the outcome model coefficients, $\boldsymbol{\theta}_s$, to vary across studies. We let $\boldsymbol{\theta}_s \overset{iid}{\sim} N(\boldsymbol{\mu}, \boldsymbol{\Sigma}_{\theta})$, with $\boldsymbol{\mu} = \boldsymbol{\theta}_0$ and 
\begin{equation*}
    \boldsymbol{\Sigma}_{\theta} = \begin{pmatrix}
    0.5 & 0 & 0 & 0\\
    0 & \Sigma_{\gamma} & 0 & 0 \\
    0 & 0 & 0.5 & 0 \\
    0 & 0 & 0 & \Sigma_{\kappa} \\
    \end{pmatrix}.
\end{equation*}
For all $s \in \mathcal{S}$, the $\boldsymbol{\theta}_s$ were mutually independent of $X_i$, $A_i$, $S_i$, and $\epsilon_i$ for all $i=1,2,...,n$. 

We focused on considering performance when the $\kappa_s$ vary more than the $\gamma_s$ (so the between-study treatment effect heterogeneity in the conditional ATEs is driven by differences in the impact of the measured features) and when the $\gamma_s$ vary more than the $\kappa_s$ (so the between-study treatment effect heterogeneity in the conditional ATEs is driven by differences in the impact of unmeasured features associated with study). For the former, we set $\Sigma_{\gamma} =0.1$ and $\Sigma_{\kappa} = 2$. For the latter, we set $\Sigma_{\gamma} =2$ and $\Sigma_{\kappa} = 0.1$. 

The assumptions made in developing our proposed two-stage estimator are met in these settings, because we have set the mean of the distribution of the $\boldsymbol{\theta}_s$ to be the vector of coefficients for the outcome generation model in the target population. In contrast, the assumptions used to develop the pooled estimator are not met. We chose to focus on these settings where the coefficients in the outcome generation model are random across studies because we believe this to be more realistic than assuming them to be fixed effects. 

\subsection{Estimation and Evaluation}

For all three settings, we studied the performance of three estimators: (1) the na\"{\i}ve unadjusted estimator implemented as in Equation \ref{Eq: Pooled_WtEst} with weights as in Equation \ref{Eq: SingleStudy_SimpleWts}, (2) the pooled estimator implemented as in Equation \ref{Eq: Pooled_WtEst} with weights as in Equation \ref{Eq: MultStudy_PooledWts}, and (3) our proposed two-stage estimator implemented as in Equation \ref{Eq: TwoStage_Est} with even study weights, such that $w_s = 1$ for $s \in \mathcal{S}$. We evaluate estimator performance via bias, mean squared error (MSE), and empirical standard error (EmpSE). Summarizing over the estimates for the simulation replicates, we estimated the bias with the estimates' average distance from the true TATE, the MSE with the average squared distance from the true TATE, and the EmpSE with the square root of their sample variance. To calculate the true TATE in each setting, we obtained empirical approximations of the expected value of $X$ in the target population from a simulation with 50,000 replications. The estimated expected values in the target population, along with their Monte Carlo standard errors, can be found in Appendix C. 

In addition to studying estimator performance, we examine the performance of the stratified bootstrap procedure we proposed in Section \ref{SubSec: Var_Est} in estimating the variance of the estimators and constructing confidence intervals. For each simulation replication, we took 1,000 bootstrap samples. We estimated standard errors by taking the square root of the sample variance of the bootstrapped estimates. Approximate 95\% confidence intervals were calculated based on bootstrap percentiles. We evaluate the performance of the bootstrap procedure by considering the coverage of the percentile-based 95\% confidence intervals and the relative error in the bootstrapped standard error estimates. The latter were calculated as $100\left(\frac{\widehat{\text{Avg. BS-SE}}}{\widehat{\text{Emp. SE}}}-1\right)\%$, where $\widehat{\text{Avg. BS-SE}}$ is the average of the estimated standard errors obtained from the bootstrap procedure over the simulation replications and $\widehat{\text{Emp. SE}}$ is the estimated empirical standard error. For computational expediency, we focus only on setting 2 and use only 500 replications for the m=30 setting.

Simulations were conducted using R version 3.5.1 \cite{rcoreteam}. Logistic regression models were fit using the glm function in the stats base R package. Multinomial logistic regression models were fit using the vglm function in the VGAM package \cite{yee2015vector,yee1996vector}, where estimate convergence was assessed directly on the key feature for our method -- the model coefficient estimate values. Figures and summaries were produced using R version 4.2.1 \cite{rcoreteamv421} and the package rsimsum \cite{gasparini2018rsimsum} was used to estimate bias, MSE, and EmpSE.

\begin{table}[!p]
    \centering
    \begin{tabular}{ccccc}
    \multicolumn{5}{c}{Three Studies} \\
    & Estimator & Setting 1 & Setting 2 & Setting 3 \\
    \hline 
    \hline 
     \multirow{3}{*}{Bias} & Unadjusted & 0.091 (0.0080) & 0.056 (0.0074) & 0.099 (0.0286) \\
        \cline{2-5}
        & Pooled & \textbf{-0.007 (0.0077)} & \textbf{-0.002 (0.0079)} & \textbf{0.040 (0.0287)} \\
        \cline{2-5}
        & Two-Stage & \textbf{-0.003 (0.0062)} & \textbf{-0.006 (0.0061)} & \textbf{0.035 (0.0256)} \\
        \hline \hline
        \multirow{3}{*}{EmpSE} & Unadjusted & 0.252 (0.0056) & 0.235 (0.0053) & 0.903 (0.0202) \\
        \cline{2-5}
        & Pooled & 0.242 (0.0054) & 0.249 (0.0056) & 0.906 (0.0203) \\
        \cline{2-5} 
        & Two-Stage & \textbf{0.195 (0.0044)} & \textbf{0.194 (0.0043)} & \textbf{0.809 (0.0181)} \\
        \hline \hline
        \multirow{3}{*}{MSE} & Unadjusted & 0.071 (0.0032) & 0.058 (0.0025) & 0.824 (0.0359) \\
        \cline{2-5}
        & Pooled & 0.059 (0.0025) & 0.062 (0.0028) & 0.822 (0.036) \\
        \cline{2-5}
        & Two-Stage & \textbf{0.038 (0.0016)} & \textbf{0.038 (0.0017)} & \textbf{0.655 (0.0307)} \\
        \\
        \multicolumn{5}{c}{Thirty Studies} \\
        & Estimator & Setting 1 & Setting 2 & Setting 3 \\
    \hline 
    \hline 
        \multirow{3}{*}{Bias} & Unadjusted & 0.094 (0.0021) & 0.058 (0.0023) & 0.053 (0.0094) \\
        \cline{2-5}
        & Pooled & \textbf{0.002 (0.0025)} & \textbf{-0.003 (0.0027)} & \textbf{-0.007 (0.0095)} \\
        \cline{2-5}
        & Two-Stage & \textbf{0.003 (0.0023)} & \textbf{-0.004 (0.0021)} & \textbf{-0.008 (0.0079)} \\
        \hline \hline
       \multirow{3}{*}{Emp. SE} & Unadjusted & 0.066 (0.0015) & 0.073 (0.0016) & 0.298 (0.0067) \\
        \cline{2-5}
        & Pooled & 0.079 (0.0018) & 0.085 (0.0019) & 0.299 (0.0067) \\
        \cline{2-5}
        & Two-Stage & \textbf{0.073 (0.0016)} & \textbf{0.067 (0.0015)} & \textbf{0.250 (0.0056)} \\
        \hline \hline
        \multirow{3}{*}{MSE} & Unadjusted & 0.013 (0.0004) & 0.009 (0.0004) & 0.092 (0.0039) \\
        \cline{2-5}
        & Pooled & 0.006 (0.0003) & 0.007 (0.0003) & 0.089 (0.0038) \\
        \cline{2-5} 
        & Two-Stage & \textbf{0.005 (0.0002)} & \textbf{0.005 (0.0002)} & \textbf{0.063 (0.0029)} \\
    \end{tabular}
    \caption{Results from simulations with three and thirty studies; simulations had 1,000 replications; Monte Carlo standard errors are shown in parentheses; lowest values in for each criteria for each setting are highlighted with bold font}
    \label{tab:simresults}
\end{table}

\subsection{Simulation Results}

Results from simulations with three and thirty studies can be found in Table \ref{tab:simresults}. Not surprisingly, the unadjusted estimator is biased towards the treatment effect in the largest study, and this bias dominates the MSE. In contrast, both the pooled and two-stage estimators are essentially unbiased in all settings considered, leading to the EmpSE essentially determining the MSE. Under all settings considered, the two-stage estimator had lower EmpSE than the existing pooled estimator. The lower variability for the two-stage estimator is most pronounced in setting 3, where the study sizes differed and unmeasured study features were the main source of differences in treatment effect heterogeneity across studies.

The 95\% confidence interval coverage and the relative error in the bootstrapped standard error estimates can be found in Table \ref{tab:cov_simresults}. Corresponding results tables can be found in Appendix \ref{sims.app}. We can see in Table \ref{tab:cov_simresults} that when simulations contained three studies, the bootstrap procedure underestimated estimators' standard errors and produced confidence intervals with less than desired coverage. These results reflect that it is difficult to estimate between-study variation when only a small number of studies are available. When simulations instead contained thirty studies, we see that our proposed stratified bootstrap procedure has better performance with the coverage of the percentile-based 95\% confidence intervals for the pooled and two-stage estimators much closer to desired levels than in the three study simulations.

\begin{table}[t]
    \centering
    \begin{tabular}{c||c||c|c}
    \hline 
    \hline 
    & Estimator & Three Studies & Thirty Studies \\
    \hline 
    \hline 
     \multirow{3}{*}{Coverage} & Unadjusted &  0.772 (0.0133) & 0.826 (0.017) \\
        \cline{2-4}
        & Pooled & 0.813 (0.0123) & 0.970 (0.0076) \\
        \cline{2-4}
        & Two-Stage & 0.761 (0.0135) & 0.962 (0.0086) \\
        \hline
        \multirow{3}{*}{Rel. Error} & Unadjusted &  -5.16\% (2.97) & 17.9\% (3.81)\\
        \cline{2-4}
        & Pooled & -28.3\% (2.07) & 5.30\% (3.40) \\
        \cline{2-4} 
        & Two-Stage & -25.3\% (2.22) & 7.20\% (3.44) \\
    \hline 
    \hline 
    \end{tabular}
    \caption{Coverage of the 95\% percentile intervals and the relative error in the bootstrapped standard error estimates derived using proposed stratified bootstrap procedure for simulations under Setting 2; Simulations with $m=3$ studies had 1,000 replications; Simulations with $m=30$ studies had 500 replications; Bootstrapped estimates were derived using 1,000 bootstrap samples; Monte Carlo standard errors are shown in parentheses}
    \label{tab:cov_simresults}
\end{table}

\section{Application to TBI} \label{Sec: TBI}

\begin{table}[t]
\centering
\begin{tabular}{l|c|c|c|c}
    \hline
    \multirow{2}{*}{Study} & Clinical Trial & \multirow{2}{*}{Control} & \multirow{2}{*}{Age$^{a}$} & Time Since  \\
     & Number & & & Injury$^{b}$  \\
    \hline
    \hline 
    CDC \cite{wade2006online}  &  NCT00178022 & Usual Care & 5-18 & 0-24 \\
    Online R21 \cite{wade2017teen} & $^{c}$ & IRC & 5-18 & 0-24\\ 
    Original TOPS \cite{wade2011effect} & NCT00409058 & IRC & 11-18 & 0-24 \\
    TOPS-RRTC \cite{narad2015effects} & NCT01042899 & IRC & 11-18 & 0-18 \\
    \hline 
    \multicolumn{5}{l}{IRC: Internet resource comparison}\\
    \multicolumn{5}{l}{$^{a}$in years; $^{b}$in months}\\
    \multicolumn{5}{l}{$^{c}$Study predated clinicaltrials.gov}\\
\end{tabular}
\caption{Trial and recruitment summaries for studies included in our application to studies on pediatric TBI}
\label{Tab: TBI_RecruitmentSum}
\end{table}

Here, we illustrate our proposed two-stage approach to extending inferences using our example application to a meta-analysis of a collection of four RCTs that studied the effect of OFPST on symptoms in children who had been hospitalized for moderate-to-severe TBI \cite{narad2015effects, wade2018online, wade2017teen, wade2011effect}. Each study randomly assigned children, along with their caregivers, to treatment with OFPST plus access to internet resources or to control, which was access to internet resources only or usual care. Enrollment criteria for each study were based on the child's age, in years, and their time since injury, in months. More information about each study and its recruitment criteria can be found in Table \ref{Tab: TBI_RecruitmentSum}.

We consider a target population of children aged 11 to 13 years who had experienced a moderate-to-severe TBI within the last three months. A sample from the target population was obtained from a cohort study on children between the ages of 6 and 13 who had experienced a moderate-to-severe TBI \cite{wade2002prospective}. This cohort study followed children for up to four years, with follow-ups at six months, one year, and four years. For this analysis, we consider the subset of the cohort study sample at baseline of children aged 11 to 13 years who had experienced a moderate-to-severe TBI within the last three months to be an SRS from the target population. 

\begin{table}[t]
\centering
\begin{tabular}{lccc}
  \hline
  & Target & CDC & Online R21 \\
 & ($n=55$) & ($n=39$) & ($n=34$) \\
  \hline
    \hline
  EPS & 51.6 (10.3) & 55.1 (12.8) & 55.0 (12.5) \\
  Age$^{a}$ & 12.0 (0.546) & 12.0 (3.52) & 11.3 (3.18) \\
  TSI$^{a}$ & 0.0603 (0.0314) & 0.346 (0.270) & 1.09 (0.601) \\
  Male &  41 (74.5\%) & 22 (56.4\%) & 19 (55.9\%) \\
  Severe TBI & 15 (27.3\%) & 12 (30.8\%) & 11 (32.4\%) \\
   \hline
 & Original TOPS & TOPS-RRTC &\\ 
 & ($n=35$) & ($n=72$) & \\ 
  \hline
    \hline
  EPS & 53.4 (10.6) & 51.9 (11.9)& \\         
  Age$^{a}$ & 14.3 (2.35) & 14.8 (2.03) & \\ 
  TSI$^{a}$ & 0.807 (0.423) & 0.490 (0.347) & \\ 
  Male & 17 (48.6\%) & 51 (70.8\%) & \\ 
  Severe TBI & 14 (40.0\%) & 29 (40.3\%) & \\ 
   \hline
   \multicolumn{4}{l}{EPS: Externalizing problems score}\\
    \multicolumn{4}{l}{TSI: Time since injury}\\
 \multicolumn{4}{l}{$^{a}$in years.}\\
\end{tabular}
\caption{TBI Baseline Covariate Summaries; categorical summaries shown as Count (\%); continuous summaries shown as Mean (SD)} 
\label{Tab: TBI_Summary}
\end{table}

We focused on estimating the TATE of six months of OFPST on the Child Behavior Checklist's (CBCL) externalizing problems score. To account for baseline differences in problems scores, we used change in score from baseline to six months as our outcome because an individual's treatment effect for change in their score, $(Y^a - Y_{baseline}) - (Y^{a'} - Y_{baseline})$, is the same as their treatment effect if we just considered score, $Y^{a}-Y^{a'}$. CBCL externalizing problems scores were examined as t-scores on the Externalizing Problems scale where smaller scores indicate more well-adjusted externalizing behaviors. Therefore, a negative treatment effect would indicate that OFPST is beneficial. We considered the following baseline covariates: age, time since injury, sex, and severity of TBI. For ease, we only considered target observations with complete baseline covariate data and study observations with complete outcome and baseline covariate data. This resulted in an overall sample size of $n=241$ with 55 observations in the target population sample. Baseline covariate summaries by study, including baseline externalizing problems score, can be found in Table \ref{Tab: TBI_Summary}. 

\subsection{Study-Specific Estimates}

\begin{figure}[t]
    \centering
    \includegraphics[width=\textwidth]{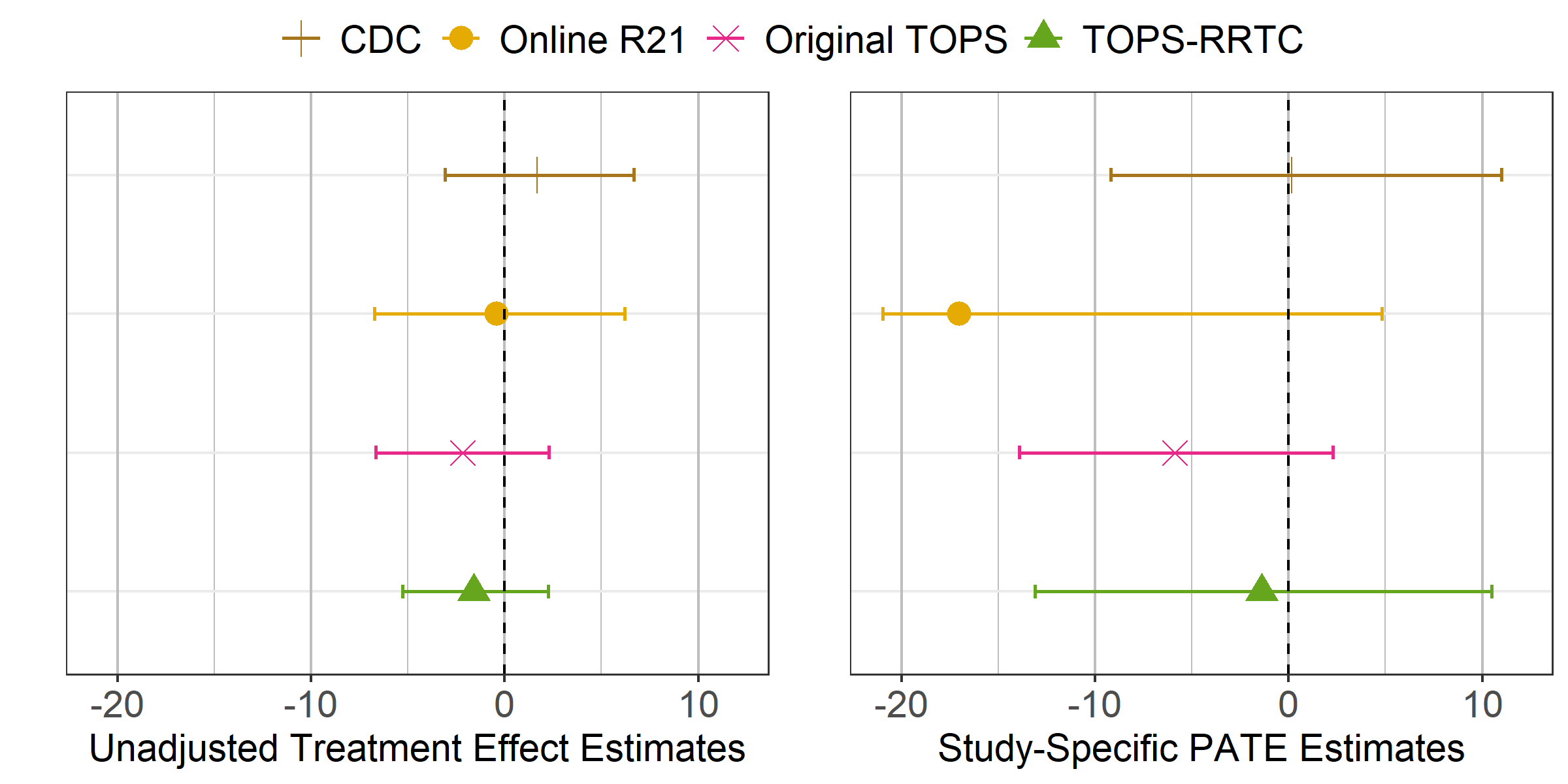}
    \caption{Unadjusted treatment effect estimates and study-specific TATE estimates from each of the four pediatric TBI studies in our collection; plotted intervals are 95\% bootstrapped percentile intervals}
    \label{Plot: StudySpecEsts_TBI}
\end{figure}

To consider why we want to utilize methods for extending inferences from a collection of studies in this setting, we first considered unadjusted treatment effect estimates and study-specific TATE estimates from each of the four studies include in our analysis. The resulting estimates are plotted in Figure \ref{Plot: StudySpecEsts_TBI} along with 95\% bootstrapped confidence intervals. The unadjusted estimates are simply the difference between the average outcome among the treated and the average outcome among the control within a study's sample. The study-specific TATE estimates were obtained using the an existing method for extending inferences from single study that takes the same form as the pooled and two-stage estimator when used to extend from a single study to extend inferences from each study to the target population. 

In Figure \ref{Plot: StudySpecEsts_TBI}, we see that the unadjusted treatment effect estimates from each study are fairly similar while the study-specific TATEs have larger differences across study, though their 95\% bootstrapped percentile intervals overlap. Recall that each study considered in this analysis was a clinical trial considering a different population of interest. If we were to obtain an overall estimate by pooling these unadjusted estimates, as would typically be done in an traditional two-stage meta-analysis, it would not be clear what population the resulting estimate relates to. When we instead adjust each study to represent the target population, we see that resulting estimates have more heterogeneity than may have been expected based solely on the unadjusted treatment effect estimates. This suggests that we may be concerned about across-study heterogeneity in treatment effects. 

\subsection{Considering Assumptions}

\begin{table}[t]
\centering
\begin{tabular}{lccc}
  \hline
  & Target & CDC & Online R21 \\
 & ($n=55$) & ($n=45$) & ($n=34$) \\
  \hline
    \hline
  Age$^{a}$ & 12.0 [11.0, 13.0] & 12.0 [5.31, 17.4] & 11.3 [6.47, 18.0] \\ 
  TSI$^{a}$ & 0.0603 [0.0177, 0.156] & 0.359 [0.0600, 0.950] & 1.09 [0.150, 2.26] \\ 
   \hline
 & Original TOPS & TOPS-RRTC &\\ 
 & ($n=35$) & ($n=72$) & \\ 
  \hline
    \hline
  Age$^{a}$ & 14.3 [11.1, 18.4]  & 14.8 [11.3, 18.5] \\ 
  TSI$^{a}$ & 0.807 [0.240, 1.68] & 0.490 [0.0600, 1.51] \\ 
   \hline
   \multicolumn{4}{l}{TSI: Time since injury; $^{a}$in years.}\\
\end{tabular}
\caption{Summaries of eligibility criteria for the TBI studies and the sample form the target population shown as Mean [Min, Max]}
\label{Tab: Recruit_Sums}
\end{table}

All of the studies included in our collection are RCTs, so it is reasonable to expect that Assumptions A2 and A3, which respectively assume the exchangeability of potential outcomes and positivity of treatment assignment hold. Because the studies in our collection were not part of one larger trial but were instead conducted by different (though overlapping) research groups at different times, utilized different enrollment sites, and employed slightly different eligibility criteria, we may expect that an individual's potential outcome under a given treatment and/or their ATE, conditional on their covariates, depends on which study they were enrolled in. If this is true, assumptions A1 and A4, required for the identifiability and consistency of the pooled estimator, are not reasonable in this setting. Assumption B1, which allows for setting where an observation's potential outcome changes if their study membership changes, and Assumption B4, which allows for between-study heterogeneity in conditional ATEs, may still be reasonable so long as we are willing to assume that the conditional ATE in the target population is the expected value of the conditional ATEs in the studies. 

Because of how we have chosen our target population, we know that Assumptions A5, which assumes positivity of participation in any study for each observation in the target population, are reasonable in theory. To check for practical positivity issues, we estimated the probability of each observation in the sample from the target population participating in each study by fitting separate logistic regression models for study-participation within each study, using both the study and target data. First, we considered models that included main effects for each of the baseline covariates as well as a quadratic term for age because the age range in our target population falls in the middle of the range considered in two of the studies and on the low end of the range in the other two studies. This approach lead to model fit and convergence issues. Looking at Table \ref{Tab: Recruit_Sums}, which contains summaries of the eligibility criteria, age and time since injury, by study, shows that the target sample has a much shorter range of time since injury in addition to its smaller age range. This lack of overlap across two variables caused practical positivity issues. For this exercise, we then excluded time since injury from all models moving forward, assuming that its exclusion does not violate needed identifiability assumptions. If, for example, treatment was conditionally randomized on time since injury in at least one study this exclusion would not be appropriate and we would need to consider other approaches to address the positivity issues, if possible. When models were fit without a main effect for time since injury, all observations in the target sample had a positive predicted probability of participation in each of the studies with no predicted probabilities close enough to zero to cause concern about practical positivity violations. Summaries of the predicted probabilities can be found in Table \ref{Tab: TBI_Positivity}. 

\begin{table}[t]
\centering
\begin{tabular}{l|c|c|c}
  \hline
 Study & Mean & Std. Deviation & Range \\ 
  \hline
  \hline
CDC & 0.150 & 0.104 & (0.043, 0.467) \\ 
  Online R21 & 0.094 & 0.117 & (0.020, 0.570) \\ 
  Original TOPS & 0.196 & 0.149 & (0.076, 0.727) \\ 
  TOPS-RRTC & 0.249 & 0.134 & (0.142, 0.659) \\ 
   \hline
\end{tabular}
\caption{Summaries of predicted probabilities of each member of the target population sample participating in the studies considered in the application to studies on pediatric TBI}
\label{Tab: TBI_Positivity}
\end{table}

\subsection{Results}

In order to estimate the TATE using our proposed two-stage approach, estimates of the propensity of treatment for each individual were obtained using a logistic regression model. A multinomial logistic regression model was used to obtain study-specific estimates of the odds of participation in each study as opposed to the target population. All models included main effects of age and sex, an indicator of having experienced a severe TBI, and a quadratic term for age. Variance estimates and confidence intervals were obtained using the stratified bootstrap procedure outlined in Section \ref{SubSec: Var_Est} with 10,000 replications. 

We estimated that the TATE of six months of OFPST on externalizing problems in children aged 11 to 13 years who had experienced a moderate-to-severe TBI within the last three months is -4.52 with a 95\% percentile interval of (-11.98, 2.78) when compared to access to internet resources only or usual care. Given that our sample size and target population were constrained by the need for positivity, it is not surprising that we did not obtain a statistically significant effect estimate, as is possible within each more powerful study. Additionally, it is important to note that these results should not be interpreted clinically without further scientific consideration of appropriate baseline covariates and model specification. 

We note that our bootstrap procedure produced 55 missing values due to issues with positivity of study membership. This is a rate of 0.55\%. To evaluate the sensitivity of our results to these missing values, we considered the percentile intervals that would result if all of the replications with missing values had estimates that were lower than all other estimates, the same as the mean estimate, or higher than all other estimates. The resulting intervals are shown in Table \ref{Tab: TBI_Interval_Sensitivity}. If the TATE estimates from the bootstrap replications that produced missing values were near the mean estimate for a given estimator, we would see minimal changes in the percentile intervals. Even if the missing TATE estimates were more extreme, all lower or higher than the other bootstrapped estimates, we do not see large differences in the percentile intervals and we observe no changes in statistical significance. 

\begin{table}[!b]
    \centering
    \begin{tabular}{c|c}
    \hline 
       Supposed Location of Missing Values  & 95\% Percentile Interval \\
       \hline 
       \hline 
         Lower than all Estimates & (-12.48, 2.78) \\
         At the Mean Estimate & (-11.97, 2.78) \\
         Higher than all Estimates & (-11.97, 3.17)\\
    \hline 
    \end{tabular}
\caption{Percentile intervals from the TBI analyses calculated considering missing values as lower than than all other estimates, to be the same as the mean estimate, and as higher than all other estimates} 
\label{Tab: TBI_Interval_Sensitivity}
\end{table}

\section{Discussion} \label{Sec: Discussion}

In this paper, we discussed methods for extending inferences from a collection of RCTs in order to obtain an estimate of the effect a treatment will cause on average in a target population of interest. Existing methods pool IPD across the studies included in the collection and assume that participation in a particular study is not associated with a participant's expected treatment effect conditional on their baseline covariates. We proposed an alternative two-stage approach that allows for between-study heterogeneity in these conditional ATEs, enabling us to extend inferences from a collection of RCTs in some settings where we expect differences in the effect of treatment across the studies included in our analyses, even after controlling for baseline covariates. Inspired by traditional methods for two-stage meta-analysis and methods for extension from a single study, our proposed approach utilizes weighting in two stages, taking a weighted average of study-specific estimates of the treatment effect in the target population.

In order to allow for extension of inferences in settings where participants' potential outcomes or conditional ATEs may depend on what study they were enrolled in, we proposed a new collection of identifiability assumptions that adaptation of the collection previously proposed for the pooled approach by Dahabreh et al. \cite{dahabreh2020toward, dahabreh2019efficient}. To obtain variance estimates and approximate confidence intervals, we proposed a stratified bootstrap procedure. We also showed that the TATE is identifiable, under our proposed collection of identifiability assumptions, as the expected value over study membership of the treatment effects that would be observed if each member of the target population was enrolled in a particular study. Additionally, we outlined a conjecture that our proposed two-stage estimator with even study-level weights is a consistent estimator of the TATE under these assumptions and a collection of regularity conditions when the number of studies grows with the number of observations at a bounded rate and the treatment and study membership models are correctly specified. 

Simulation studies showed that in at least some settings where there is between-study heterogeneity in conditional treatment effects, we can expect our proposed two-stage estimator to lower empirical standard error as compared to the existing pooled estimator. Our simulations also demonstrated that we may have issues with underestimation of variance and poor coverage of confidence intervals for both the pooled and two-stage estimators when we do not have a large collection of studies. 

To illustrate the application of the methods for extending inferences, we also considered a real data example to a collection of clinical trials studying the impact of an online therapy treatment on symptoms of pediatric TBI. This application illustrated the use of the methods we considered in a setting where between-study treatment effect heterogeneity is expected. In this example, we saw the practical positivity issues that can arise from lack of covariate overlap between the studies and the target population, even when needed theoretical positivity assumptions are reasonable. We expect these kinds of issues will arise more frequently when applying our proposed two-stage method than existing pooled methods because we require that each study can be separately used to generalize to the target population. Because existing pooled approaches are not appropriate for settings with between-study effect heterogeneity, practical positivity issues may leave us with a choice among a collection of inappropriate methods in such settings. Further research to extend the current two-stage model are needed to better utilize the full study data and more appropriately capture the full knowledge provided by the individual studies.

To allow for between-study heterogeneity in conditional treatment effects across studies, we relaxed assumptions proposed for existing approaches that pool studies' IPD. One of those original assumptions, Assumption A4, states that for each set of baseline covariates possible in the target population, the conditional ATE in each study is the same as the conditional ATE in the target population. In our proposed method, this assumption is relaxed to B4 so we instead assume that for each set of baseline covariates possible in the target population, the study-specific conditional ATEs follow some distribution with an expected value that is the conditional ATE in the target population. While we believe that our proposed assumption is more realistic than those proposed for existing methods, we acknowledge that this is still quite a strong assumption and may not always be reasonable. In practice, it would be useful to have guidelines for evaluating when utilizing a two-stage approach to extending inference improves performance of estimators enough to justify the additional complication. Developing such guidelines is beyond the scope of this paper. 

In this paper, we focused on methods for extending inferences that involve taking weighted averages of observed outcomes, which we refer to as weighted methods. These methods are appealing, in part, because they do not require us to assume that we can develop an appropriate model for the outcome, a task that could be challenging even for subject-matter experts. If we believe that we are able to construct a reasonable outcome model, another approach to extending inferences would involve modeling the conditional expectation of the potential outcomes in the target population. Such approaches have been proposed for extending inferences from a single study (e.g, \cite{buchanan2018generalizing,dahabreh2019extending, dahabreh2020toward}). For extension from a collection of studies, approaches that rely on pooling the study data have also been proposed and, like the pooled weighted estimator we considered in this work, assume that an observation's expected treatment effect conditional on their covariates does not depend on what study they were enrolled in (e.g, \cite{dahabreh2020toward, dahabreh2019efficient, steingrimsson2022systematically}). We believe that ideas used in this work could be used in concert with our two-stage framework to develop a model-based approach that allows for between-study heterogeneity in observations' conditional expected treatment effects. 

\newpage

\bibliographystyle{plain}
\bibliography{methods_ref.bib}

\newpage

\appendix

\section{Appendix A: Identifiability of the TATE}
\label{identifiability.app}

In this appendix, we prove the theorem introduced in Section \ref{Subsec: TS_ID} that states that the TATE is identifiable under the collection of identifiability assumptions we proposed in Section \ref{Subsec: TS_Assumption} and a regularity condition. 

\subsection{Set-Up}

Consider a collection of RCTs $\mathcal{S} = \{1,2,3,...,m\}$ that each consider a collection of treatments $\mathcal{A}$, measure outcome $Y$, and observe a vector of baseline covariates $\boldsymbol{X}$. Additionally, consider a non-finite target population of interest that does not include any study observations. Let $S \in \{0,\mathcal{S}\}$ be a random variable for study membership with $S= 0$ denoting the target population and let $A \in \mathcal{A}$ be a random variable denoting treatment. Recall the proposed collection of identifiability assumptions, denoted Assumption $\mathcal{B}$, is
\begin{itemize}[label={}]
    \item \textbf{B1}. For all $s \in \mathcal{S}$, $a \in \mathcal{A}$, and $i \in \{1,2,...,n:S_i=s\}$, if $A_i=a$ then $Y_i^a = Y_i$. 
    \item \textbf{A2}. For all $s\in \mathcal{S}$, for all $a \in \mathcal{A}$, and for $\boldsymbol{x}$ such that $f(\boldsymbol{x},S=s) > 0$, 
    \begin{equation*}
        E(Y^a| \boldsymbol{X}=\boldsymbol{x},S=s, A=a) = E(Y^a|\boldsymbol{X}=\boldsymbol{x},S=s).
    \end{equation*}
    \item \textbf{A3}. For all $s\in \mathcal{S}$, for all $a \in \mathcal{A}$, and for $\boldsymbol{x}$ such that $f(\boldsymbol{x},S=s) > 0$,
    \begin{equation*}
        P(A=a|\boldsymbol{X}=\boldsymbol{x},S=s) >0.
    \end{equation*}
    \item \textbf{B4}. For all $a, a' \in \mathcal{A}$ and for $\boldsymbol{x}$ such that $f(\boldsymbol{x},S=0) > 0$, 
    \begin{equation*}
        E(Y^a - Y^{a'}|\boldsymbol{X}=\boldsymbol{x},S=0) = E_{S}\left[E(Y^a - Y^{a'}|\boldsymbol{X}=\boldsymbol{x},S)\big|S \in \mathcal{S} \right].
    \end{equation*}
    \item \textbf{A5}. For all $s\in \mathcal{S}$ and for $\boldsymbol{x}$ such that $f(\boldsymbol{x},S=0) > 0$, 
    \begin{equation*}
        P(S=s|\boldsymbol{X}=\boldsymbol{x})>0.
    \end{equation*}
\end{itemize}
where $f(\cdot)$ is used to denote densities. 

\subsection{Main Result} 

\noindent \textbf{Theorem}: \emph{Under regularity condition $$E_S\left\{E_X\left[\left|E_{Y^a,Y^{a'}}\left(Y^a - Y^{a'}\big|\boldsymbol{X},S\right)\right|\big|S=0 \right]\big|S \in \mathcal{S} \right\} < \infty$$ and Assumption $\mathcal{B}$}, 
\begin{equation}  \label{Eq: IdentfiabilityResult1}
\begin{split}
    \Delta \equiv E(Y^a-Y^{a'}|S=0) = E_S\left(\Delta_S|S \in \mathcal{S}\right)
\end{split}    
\end{equation}
\emph{with} 
\begin{equation} \label{Eq: IdentfiabilityResult2}
\begin{split}
     \Delta_s \equiv & E_X\left[E_{Y^a,Y^{a'}}(Y^a-Y^{a'}|\boldsymbol{X},S=s)\big|S=0\right] \\
     = & E_{A,X,Y}\left(\left[\frac{w(a,s)}{E_{A,X}\left(w(a,s)|S=s\right)} - \frac{w(a',s)}{E_{A,X}(w(a',s)|S=s)}\right]Y\Big|S=s\right)
\end{split}
\end{equation}
\emph{for each $s \in \mathcal{S}$ where
\begin{equation*}
    w(a,s) = \frac{I(A=a)}{P(A=a|\boldsymbol{X}, S=s)}\frac{P(S=0|\boldsymbol{X})}{P(S=s|\boldsymbol{X})}.
\end{equation*}
for each $s \in \mathcal{S}$ and $a \in \mathcal{A}$.}

\noindent \underline{Proof of \ref{Eq: IdentfiabilityResult1}}: Let $\Delta(\boldsymbol{X},S) = E_{Y^a,Y^{a'}}\left(Y^a - Y^{a'}\big|\boldsymbol{X},S\right)$. Then we have that 
\begin{equation*}
\begin{split}
    \Delta = & E_{Y^a,Y^{a'}}\left(Y^a - Y^{a'}\big|S=0\right) \\
    = & E_X\left\{E_{Y^a,Y^{a'}}\left[Y^a-Y^{a'}\big|\boldsymbol{X},S=0\right]\Big|S=0\right\} \\
    & \emph{by iterated expectations} \\
    = & E_X\left\{E_S\left[\Delta(\boldsymbol{X},S)\big|S \in \mathcal{S} \right]\Big|S=0\right\} \\
    & \emph{ by Assumptions B4 and A5} \\
    = & E_S\left\{E_X\left[\Delta(\boldsymbol{X},S)\big|S=0 \right]\big|S \in \mathcal{S} \right\} \\
    & \emph{by Fubini's Theorem and the regularity condition} \\
    = & E_S\left(\Delta_S|S \in \mathcal{S}\right)
\end{split}
\end{equation*}

\qed 

\noindent \underline{Proof of \ref{Eq: IdentfiabilityResult2}}: We show the proof for discrete $\boldsymbol{X}$, but analogous arguments hold for $\boldsymbol{X}$ with other distributions so long as the joint distribution of $\boldsymbol{X}$ and $S$ exists. For each $s \in \mathcal{S}$, we can write 
\begin{equation} \label{Eq: ProofA2_Eq1}
\begin{split}
     \Delta_s = &  E_X\left[E_{Y^a,Y^{a'}}(Y^a-Y^{a'}|\boldsymbol{X},S=s)\big|S=0\right] \\ 
     = & E_X\left\{E_{Y^a}(Y^a|\boldsymbol{X}, S=s,A=a) |S=0\right\} \\
    & - E_X\left\{E_{Y^{a'}}(Y^{a'}|\boldsymbol{X}, S=s,A=a') |S=0\right\} \\
    & \emph{by Assumptions A2 and A3} \\
     = & E_X\left\{E_{Y}(Y|\boldsymbol{X}, S=s,A=a) |S=0\right\} \\
     & - E_X\left\{E_{Y}(Y|\boldsymbol{X}, S=s,A=a') |S=0\right\}\\
     & \emph{by Assumption B1} \\
\end{split}
\end{equation} 
Notice that, for all $a \in \mathcal{A}$ and $s \in \mathcal{S}$,
\begin{equation} \label{Eq: ProofA2_Eq2}
\begin{split}
    E_{X,A,Y}&\left[w(a,s) Y |S=s\right] = E_{X,A,Y} \left[ \frac{I(A=a)}{P(A=a|\boldsymbol{X},S)} \frac{P(S=0|\boldsymbol{X})}{P(S=s|\boldsymbol{X})} Y\Big|S=s\right] \\
    = & E_X\left\{E_{A,Y} \left[ \frac{I(A=a)}{P(A=a|\boldsymbol{X},S=s)} \frac{P(S=0|\boldsymbol{X})}{P(S=s|\boldsymbol{X})} Y\Big|\boldsymbol{X},S=s\right]\Big|S=s\right\} \\
    & \emph{by the definition of conditional expectation} \\
    = &  E_X\left\{\frac{E_{A,Y}\left[ I(A=a) Y|S=s,\boldsymbol{X}\right]}{P(A=a|\boldsymbol{X},S=s)}\frac{P(S=0|\boldsymbol{X})}{P(S=s|\boldsymbol{X})}|S=s\right\}.\\
\end{split}
\end{equation}
Additionally, we can write 
\begin{equation}  \label{Eq: ProofA2_Eq3}
\begin{split}
E_{A,Y}\big[ I(A=a) & Y\Big|\boldsymbol{X}, S=s\big] = E_A \Big( I(A=a)E_{Y}\left[ Y\Big|\boldsymbol{X}, S=s, A\right]|\boldsymbol{X}, S=s\Big)\\
    & \emph{by the definition of conditional expectation} \\
    = & \sum_{a^*\in \mathcal{A}} I(a^*=a)E_{Y}\left[ Y\Big|\boldsymbol{X}, S=s,A=a^*\right]P(A=a^*|\boldsymbol{X}, S=s)\\
    & \emph{because $A$ is discrete}\\
    = & E_{Y}\left[ Y\Big|\boldsymbol{X}, S=s, A=a\right]P(A=a|\boldsymbol{X}, S=s)\\
\end{split}
\end{equation}
because $I(a^*=a) = 0$ for all $a^* \ne a$. Equations \ref{Eq: ProofA2_Eq2} and \ref{Eq: ProofA2_Eq3} together lead to 
\begin{equation} \label{Eq: ProofA2_Eq4}
\begin{split}
    E_{X,A,Y}&\left[w(a,s) Y |S=s\right] \\
    = & \sum_{\boldsymbol{x}} \frac{P(S=0|\boldsymbol{X}=\boldsymbol{x})}{P(S=s|\boldsymbol{X}=\boldsymbol{x})}E_{Y}\left[ Y|\boldsymbol{X}=\boldsymbol{x},S=s,A=a\right] P(\boldsymbol{X}=\boldsymbol{x}|S=s) \\
    & \emph{ assuming } \boldsymbol{X} \emph{ is discrete} \\
    = & \frac{P(S=0)}{P(S=s)}\sum_{\boldsymbol{x}} E_{Y}\left[ Y|\boldsymbol{X}=\boldsymbol{x}, S=s,A=a\right]P(\boldsymbol{X}=\boldsymbol{x}|S=0) \\ 
    & \emph{by Bayes' Theorem}\\
    = &  \frac{P(S=0)}{P(S=s)}E_X\left\{E_{Y}\left[ Y|\boldsymbol{X}, S=s,A=a\right]\big|S=0\right\}
\end{split}
\end{equation}
So, from Equations \ref{Eq: ProofA2_Eq1} and \ref{Eq: ProofA2_Eq4} we have that 
\begin{equation} \label{Eq: ProofA2_Eq5}
\begin{split}
     \Delta_s = & \frac{P(S=s)}{P(S=0)}E_{A,X, Y}\left[w(a,s)Y|S=s\right] - \frac{P(S=s)}{P(S=0)}E_{A,X, Y}\left[w(a',s)Y|S=s\right]\\
\end{split}
\end{equation}
Next, notice that for all $a \in \mathcal{A}$ and $s \in \mathcal{S}$,
\begin{equation} \label{Eq: ProofA2_Eq6}
\begin{split}
    E_{A,X}&\left(w(a,s)|S=s\right)  =  E_{A,X}\left(\frac{I(A=a)}{P(A=a|\boldsymbol{X},S=s)}\frac{P(S=0|\boldsymbol{X})}{P(S=s|\boldsymbol{X})}\big|S=s\right) \\ 
    = & E_X\left[E_{A}\left(\frac{I(A=a)}{P(A=a|\boldsymbol{X},S=s)}\frac{P(S=0|\boldsymbol{X})}{P(S=s|\boldsymbol{X})}\Big|\boldsymbol{X},S=s\right)\Big|S=s\right] \\
     & \emph{by the definition of conditional expectations} \\
    = & E_X\left[\frac{1}{P(A=a|\boldsymbol{X}, S=s)}E_{A}\left(I(A=a)\Big|\boldsymbol{X}, S=s\right)\frac{P(S=0|\boldsymbol{X})}{P(S=s|\boldsymbol{X})}\Big|S=s\right] \\
    = & E_X\left[\frac{P(S=0|\boldsymbol{X})}{P(S=s|\boldsymbol{X})}\Big|S=s\right] \\
    = & \sum_{\boldsymbol{x}} \frac{P(S=0|\boldsymbol{X} = \boldsymbol{x})}{P(S=s|\boldsymbol{X}=\boldsymbol{x})}P(\boldsymbol{X}=\boldsymbol{x}|S=s)\\
    & \emph{ assuming } \boldsymbol{X} \emph{ is discrete} \\
    = & \frac{P(S=0)}{P(S=s)}\sum_{\boldsymbol{x}} P(\boldsymbol{X}=\boldsymbol{x}|S=0)\\
    & \emph{by Bayes' Theorem} \\
    = & \frac{P(S=0)}{P(S=s)}
\end{split}
\end{equation}
Finally, by combining Equations \ref{Eq: ProofA2_Eq5} and \ref{Eq: ProofA2_Eq6} we have that
\begin{equation*}
\begin{split}
    \Delta_s = & \frac{E_{A,X,Y}\left(w(a,s)Y|S=s\right)}{E_{A,X}\left(w(a,s)|S=s\right)} - \frac{E_{A,X,Y}\left(w(a',s)Y|S=s\right)}{ E_{A,X}\left(w(a',s)|S=s\right)}\\
    = & E_{A,X,Y}\left(\left[\frac{w(a,s)}{E_{X,A}\left(w(a,s)|S=s\right)} - \frac{w(a',s)}{E_{X,A}(w(a',s)|S=s)}\right]Y\Big|S=s\right)\\
\end{split}
\end{equation*} 

\qed

\newpage 

\section{Appendix B: Consistency of the Two-Stage Weighted Estimator of the TATE}
\label{consistency.app} 

We conjecture that our proposed two-stage estimator of the TATE with even study-level weights is a consistent estimator under the identifiability assumptions we presented in Section \ref{Subsec: TS_Assumption} and some regularity conditions if the number of studies is growing with the sample size at a bounded rate and the treatment and study membership models are correctly specified. This appendix contains a detailed outline of our reasoning for logistic treatment models and multinomial logistic study membership models. 

\subsection{Notation and Assumptions} \label{Const: NandA}

Suppose that we have access to data on $n$ observations from a collection of $m$ RCTs, $\mathcal{S} = \{1,2,...,m\}$, and an SRS from a target population. Let $Y_i$, $i=1,2,...,n$, be the outcome of interest, $\boldsymbol{X}_i$ be a length $q-1$ vector of baseline covariates, and $A_i \in \mathcal{A}$ be the treatment where $\mathcal{A}$ is a finite set. Let $S_i \in \{0,\mathcal{S}\}$ indicate study membership where $S_i=0$ if observation $i$ is in the target population. Assume the following regularity conditions: 
\begin{itemize}[label={}]
    \item \textbf{(i)} From each $s \in \mathcal{S}$ we have $n_s$ observations $\{\boldsymbol{X}_i, S_i=s, A_i,Y_i\}$ that are independent and identically distributed for $i \in \{1,...,n: S_i =s\}$ and $n = \sum_{s \in \mathcal{S}} n_s$.
    \item \textbf{(ii)} From the target population we have $n_0$ observations $\{\boldsymbol{X}_i,S_i=0\}$ that are independent and identically distributed for $i \in \{1,...,n: S_i =0\}$.
    \item \textbf{(iii)}  For each $s,k \in \mathcal{S}$ with $s \ne k$, $\{\boldsymbol{X}_i, S_i=s, A_i,Y_i\}$ is independent of $\{\boldsymbol{X}_j, S_j=k, A_j,Y_j\}$ for each $i \in \{1,...,n: S_i =s\}$ and $j \in \{1,...,n: S_j =k\}$. 
    \item \textbf{(iv)}  For all $i,j \in \{1,2,...,n\}$ with $i \ne j$, $\{\boldsymbol{X}_i, S_i\}$ is independent of $\{\boldsymbol{X}_j, S_j\}$
    \item \textbf{(v)} $E_S\left\{E_X\left[\left|E_{Y^a,Y^{a'}}\left(Y^a - Y^{a'}\big|\boldsymbol{X},S\right)\right|\big|S=0 \right]\big|S \in \mathcal{S} \right\} < \infty$.
\end{itemize}
Additionally, assume that the collection of identifiability assumptions, Assumption $\mathcal{B}$, 
\begin{itemize}[label={}]
    \item \textbf{B1}. For all $s \in \mathcal{S}$, $a \in \mathcal{A}$, and $i \in \{1,2,...,n:S_i=s\}$, if $A_i=a$ then $Y_i^a = Y_i$. 
    \item \textbf{A2}. For all $s\in \mathcal{S}$, for all $a \in \mathcal{A}$, and for $\boldsymbol{x}$ such that $f(\boldsymbol{x},S=s) > 0$, 
    \begin{equation*}
        E(Y^a| \boldsymbol{X}=\boldsymbol{x},S=s, A=a) = E(Y^a|\boldsymbol{X}=\boldsymbol{x},S=s).
    \end{equation*}
    \item \textbf{A3}. For all $s\in \mathcal{S}$, for all $a \in \mathcal{A}$, and for $\boldsymbol{x}$ such that $f(\boldsymbol{x},S=s) > 0$,
    \begin{equation*}
        P(A=a|\boldsymbol{X}=\boldsymbol{x},S=s) >0.
    \end{equation*}
    \item \textbf{B4}. For all $a, a' \in \mathcal{A}$ and for $\boldsymbol{x}$ such that $f(\boldsymbol{x},S=0) > 0$, 
    \begin{equation*}
        E(Y^a - Y^{a'}|\boldsymbol{X}=\boldsymbol{x},S=0) = E_{S}\left[E(Y^a - Y^{a'}|\boldsymbol{X}=\boldsymbol{x},S)\big|S \in \mathcal{S} \right].
    \end{equation*}
    \item \textbf{A5}. For all $s\in \mathcal{S}$ and for $\boldsymbol{x}$ such that $f(\boldsymbol{x},S=0) > 0$, 
    \begin{equation*}
        P(S=s|\boldsymbol{X}=\boldsymbol{x})>0.
    \end{equation*}
\end{itemize}
hold where $f(\cdot)$ is used to denote densities. 

With even study-level weights, our proposed two-stage estimator takes the form 
\begin{equation*}
    \hat{\Delta}_{two} = \frac{1}{m}\sum_{s \in \mathcal{S}}\hat{\Delta}_s 
\end{equation*}
where 
\begin{equation*}
    \hat{\Delta}_s = \frac{1}{\sum_{i: S_i=s} \hat{w}_i(a,s)} \sum_{i: S_i=s} \hat{w}_i(a,s)Y_i - \frac{1}{\sum_{i: S_i=s} \hat{w}_i(a',s)}\sum_{i: S_i=s} \hat{w}_i(a',s)Y_i,
\end{equation*}
for each $s \in \mathcal{S}$. For each $a \in \mathcal{A}$ and $s \in \mathcal{S}$,
\begin{equation*}
\begin{split}
    \hat{w}_i(a,s) = & \frac{I(A_i=a)}{\hat{e}_a(\boldsymbol{X}_i,s)}\frac{\hat{p}(\boldsymbol{X}_i,0)}{\hat{p}(\boldsymbol{X}_i,s)} \\
\end{split}
\end{equation*}
where 
\begin{equation*}
\begin{split}
    & e_a(\boldsymbol{X}_i,s) = P(A=a|\boldsymbol{X}_i,S=s) \\
    & p(\boldsymbol{X}_i,s) = P(S=s|\boldsymbol{X}_i) \\
    & p(\boldsymbol{X}_i,0) = P(S=0|\boldsymbol{X}_i) \\ 
\end{split}
\end{equation*}
and corresponding quantities with hats are estimates of these probabilities. 

For each $s \in \mathcal{S}$, we assume that for each $i \in\{1,2,...,n:S_i = s\}$, there exists a $\boldsymbol{\theta}_{0,s} \in \mathbb{R}^{q}$ such that 
\begin{equation} \label{ConAppend:treatment_mod}
\begin{split}
    & I(A_i=a) \overset{indp}{\sim} \text{Bernoulli}(e_a(\boldsymbol{X}_i,s)) \\
     & e_a(\boldsymbol{X}_i,s) = P(A_i=a|\boldsymbol{X}_i, S_i=s) = \text{expit}\left(\boldsymbol{V}^t_i\boldsymbol{\theta}_{0,s}\right)
\end{split}
\end{equation}
where $\boldsymbol{V}_i = (1,\boldsymbol{X}_i^t)^t \in \mathbb{R}^q$. Because the number of treatment models we fit depends on the number of studies and we would like to allow the number of studies to grow with our sample size, we reframe this collection of treatment models as one larger model with $q \times m$ parameters. If we define 
\begin{equation*}
\begin{split}
    & \boldsymbol{\theta}_0 = (\boldsymbol{\theta}_{0,1}^t,...,\boldsymbol{\theta}^t_{0,m})^t \in \mathbb{R}^{qm} \\
    & \boldsymbol{Z}_{is} = I(S_i=s)\boldsymbol{V}_i \in \mathbb{R}^{q} \\
    & \boldsymbol{Z}_i = (\boldsymbol{Z}_{i1}^t,...,\boldsymbol{Z}^t_{im})^t \in \mathbb{R}^{qm}.
\end{split}
\end{equation*}
then we can write 
\begin{equation*}
\begin{split}
    & e_a(\boldsymbol{X}_i,s) = P(A_i=a|\boldsymbol{X}_i,S_i=s) = \frac{\text{exp}\{\boldsymbol{Z}^t_i\boldsymbol{\theta}_0\}}{1+\text{exp}\{\boldsymbol{Z}^t_i\boldsymbol{\theta}_0\}}  \\
    & e_{a'}(\boldsymbol{X}_i,s) = P(A_i=a'|\boldsymbol{X}_i,S_i=s) = \frac{1}{1+\text{exp}\{\boldsymbol{Z}^t_i\boldsymbol{\theta}_0\}}  \\
\end{split}
\end{equation*} 

For each $i=1,2,..,n$, we also assume that there exists a $\boldsymbol{\beta}_{0,s} \in \mathbb{R}^q$ such that 
\begin{equation} \label{ConAppend: studymem_model}
\begin{split}
 & \left(I(S_i=0),...,I(S_i=m) \right)\overset{indp}{\sim} \text{MN}(1,p(\boldsymbol{X}_i,0),...,p(\boldsymbol{X}_i,m))\\
 & \text{log}\left(\frac{p(\boldsymbol{X}_i,s)}{p(\boldsymbol{X}_i,0)}\right) = \boldsymbol{V}^t_i\boldsymbol{\beta}_{0,s} \\
\end{split}
\end{equation}
for $s \in \mathcal{S}$ such that $\sum_{s\in \mathcal{S}} p(\boldsymbol{X}_i,s) =1$ for all $i$. (Notice that, for ease, we are assuming that the same vector of coefficients yields the true treatment and study membership models.) If we define 
\begin{equation*}
\begin{split}
    & \boldsymbol{\beta}_0 = (\boldsymbol{\beta}_{0,1}^t,...,\boldsymbol{\beta}^t_{0,m})^t \in \mathbb{R}^{mq} \\
    & \boldsymbol{U}_{is} = e_s \otimes \boldsymbol{V}_i \in \mathbb{R}^{mq}
\end{split}
\end{equation*}
where $e_s$ is the $s$th column of the $m \times m$ identity matrix and $\otimes$ denotes the Kronecker product. Then $\boldsymbol{V}_i^t \boldsymbol{\beta}_{0,s} = \boldsymbol{U}^t_{is}\boldsymbol{\beta}_0$ and  
\begin{equation*}
\begin{split}
    &p(\boldsymbol{X}_i,s)= P(S_i=s|\boldsymbol{X}_i) =  \frac{\text{exp}\{\boldsymbol{U}^t_{is}\boldsymbol{\beta}_{0}\}}{1+ \sum_{k=1}^m\text{exp}\{\boldsymbol{U}^t_{ik}\boldsymbol{\beta}_{0}\}} \\
    & p(\boldsymbol{X}_i,0) = P(S_i=0|\boldsymbol{X}_i) = \frac{1}{1+ \sum_{k=1}^m\text{exp}\{\boldsymbol{U}^t_{ik}\boldsymbol{\beta}_{0}\}}. 
\end{split}
\end{equation*}
for $s \in \mathcal{S}$.

\subsection{Consistency Conjecture}

We conjecture that our proposed estimator $\hat{\Delta}_{two}$ with even study-level weights is a consistent estimator of the TATE 
\begin{equation*}
    \Delta = E_{Y^a,Y^{a'}}(Y^a-Y^{a'}|S=0)
\end{equation*}
under the setting described in Section \ref{Const: NandA} as well as a collection of regularity conditions if the number of studies grows with the sample size at a bounded rate and the sizes of the target sample and each study sample are also growing with the overall sample size. In notation, this means that we conjecture that
\begin{equation} \label{Eq: What_we_want}
\hat{\Delta}_{two} \overset{\mathbb{P}}{\rightarrow} \Delta 
\end{equation}
as $m \to \infty$ and $\text{min}\{\tilde{\boldsymbol{n}} \} \to \infty$ where $\tilde{\boldsymbol{n}} = \{n_s\}_{s=0}^m$ if $m = o(g(n))$ for a carefully chosen function $g(\cdot)$ where $n = \sum_{s=0}^m n_s$. 

\subsection{Reasoning}

From Appendix \ref{identifiability.app}, we know that under Assumption $\mathcal{B}$ and regularity condition (v) 
\begin{equation} \label{Eq: Delta_Ident}
\begin{split}
    \Delta = E_S\left(\Delta_S|S \in \mathcal{S} \right) = E_S\left(\Delta_S|S \ne 0\right)
\end{split}    
\end{equation}
where, for each $s \in \mathcal{S}$,
\begin{equation*} 
\begin{split}
     \Delta_s = & E_{A,X,Y}\left(\left[\frac{w(a,s)}{E_{A,X}\left(w(a,s)|S=s\right)} - \frac{w(a',s)}{E_{A,X}(w(a',s)|S=s)}\right]Y\Big|S=s\right)
\end{split}
\end{equation*}
and, for each $s \in \mathcal{S}$ and $a \in \mathcal{A}$,
\begin{equation*}
    w(a,s) = \frac{I(A=a)}{P(A=a|\boldsymbol{X}, S=s)}\frac{P(S=0|\boldsymbol{X})}{P(S=s|\boldsymbol{X})}.
\end{equation*}
By Equation \ref{Eq: Delta_Ident}, if we can show that
\begin{equation} \label{Eq: step1}
   \hat{\Delta}_{two}  = \frac{1}{m} \sum_{s=1}^m \hat{\Delta}_s \overset{\mathbb{P}}{\rightarrow} E_S\left(\Delta_S|S \ne 0\right) 
\end{equation}
we will have that our two-stage estimator of the TATE with even study-level weights is consistent. 

Next, by the weak law of large numbers, we have that 
\begin{equation} \label{Eq: WLLN}
    \frac{1}{m}\sum_{s=1}^m \Delta_s \overset{\mathbb{P}}{\rightarrow} E_S\left(\Delta_S|S \ne 0\right)
\end{equation}
as $m \to \infty$ if the regularity condition
\begin{itemize}[label={}]
    \item \textbf{(vi)} $\{\Delta_s\}_{s=1}^m$ are mutually independent with $E(\Delta_s) = E(\Delta_S|S \ne 0)$ for all $s \geq 1$ and $\frac{1}{m^2} \sum_{s=1}^m \text{Var}(\Delta_s) \to 0 $ as $m \to \infty$
\end{itemize}
holds. Therefore, if we assume that (vi) holds and show that 
\begin{equation} \label{Eq: Last_hard_bit}
    \frac{1}{m} \sum_{s=1}^m \left(\hat{\Delta}_s - \Delta_s\right) \overset{\mathbb{P}}{\rightarrow} 0
\end{equation}
as $m \to \infty$ and $\text{min}\{\tilde{\boldsymbol{n}} \} \to \infty$ with $m = o(g(n))$ for a carefully chosen function $g(\cdot)$, we will have that our two-stage estimator with even study-level weights is consistent because 
\begin{equation*}
     \hat{\Delta}_{two} = \frac{1}{m} \sum_{s=1}^m \left(\hat{\Delta}_s - \Delta_s\right) + \frac{1}{m}\sum_{s=1}^m \Delta_s  
\end{equation*}

For a fixed number of studies $m$, we can show the result in Equation \ref{Eq: Last_hard_bit} as $\text{min}\{\tilde{\boldsymbol{n}} \} \to \infty$. First, notice that if the treatment model shown in Equation \ref{ConAppend:treatment_mod} and the study membership model shown in Equation \ref{ConAppend: studymem_model} are correctly specified, $\hat{\boldsymbol{\theta}}_{s}$ and $\hat{\boldsymbol{\beta}}_{s}$ are the MLEs of $\boldsymbol{\theta}_{0,s}$ and $\boldsymbol{\beta}_{0,s}$, respectively, for each $s =1,2,...,m$, and regularity conditions (i) to (iv) hold, we have that 
\begin{equation} \label{Eq: fixedm_propensityconsistency}
\begin{split}
     & \hat{e}_a(\boldsymbol{X}_i,s) = \text{expit}\left(\boldsymbol{V}^t_i\hat{\boldsymbol{\theta}}_{s}\right) \overset{\mathbb{P}}{\rightarrow} e_a(\boldsymbol{X}_i,s) = \text{expit}\left(\boldsymbol{V}^t_i\boldsymbol{\theta}_{0,s}\right) \\
    & \hat{e}_{a'}(\boldsymbol{X}_i,s) = 1 - \text{expit}\left(\boldsymbol{V}^t_i\hat{\boldsymbol{\theta}}_{s}\right) \overset{\mathbb{P}}{\rightarrow} e_{a'}(\boldsymbol{X}_i,s) = 1-\text{expit}\left(\boldsymbol{V}^t_i\boldsymbol{\theta}_{0,s}\right) \\
\end{split}
\end{equation}
as $n_s \to \infty$ for all $s =1,2,...,m$ and 
\begin{equation} \label{Eq: fixedm_studyconsistency}
\begin{split}
     & \frac{\hat{p}(\boldsymbol{X}_i,s)}{\hat{p}(\boldsymbol{X}_i,0)} = \text{exp}\{\boldsymbol{V}^t_{i}\hat{\boldsymbol{\beta}}_{s}\}\overset{\mathbb{P}}{\rightarrow} \frac{p(\boldsymbol{X}_i,s)}{p(\boldsymbol{X}_i,s)} = \text{exp}\{ \boldsymbol{V}^t_{i}\boldsymbol{\beta}_{0,s}\}\\
\end{split}
\end{equation}
as $\text{min}\{\tilde{\boldsymbol{n}} \} \to \infty$ under some usual regularity conditions. Then, it can be shown that for all $s=1,2,...,m$
\begin{equation} \label{Eq: sumws_consist}
\begin{split}
    & \frac{1}{n_s}\sum_{i: S_i =s}\hat{w}_i(a,s) \overset{\mathbb{P}}{\longrightarrow} E_{A,X}\left(w(a,s)|S=s\right) \\
    & \frac{1}{n_s}\sum_{i: S_i =s}\hat{w}_i(a,s)Y_i \overset{\mathbb{P}}{\longrightarrow} E_{A,X,Y}\left(w(a,s)Y|S=s\right)
\end{split}
\end{equation}
if $\text{min}\{\tilde{\boldsymbol{n}} \} \to \infty$. For details of this result, see Lemma 1 in Section \ref{Sec: Const_Support}. Then for each $s=1,2,...,m$,
\begin{equation} \label{Eq: Deltas_fixedm_consistency}
\begin{split}
    \hat{\Delta}_s = & \frac{\sum_{i: S_i=s} \hat{w}_i(a,s)Y_i}{\sum_{i: S_i=s} \hat{w}_i(a,s)}  - \frac{\sum_{i: S_i=s} \hat{w}_i(a',s)Y_i}{\sum_{i: S_i=s} \hat{w}_i(a',s)}\\
    \overset{\mathbb{P}}{\rightarrow} & \frac{E_{A,X,Y}(w(a,s)Y|S=s)}{E_{A,X}\left(w(a,s)|S=s\right)} - \frac{E_{A,X,Y}(w(a',s)Y|S=s)}{E_{A,X}(w(a',s)|S=s)} \\
    & \emph{ by continuous mapping theorem} \\
    & = E_{A,X,Y}\left(\left[\frac{w(a,s)}{E_{A,X}\left(w(a,s)|S=s\right)} - \frac{w(a',s)}{E_{A,X}(w(a',s)|S=s)}\right]Y\Big|S=s\right) \\
    & = \Delta_s 
\end{split}
\end{equation}
as $\text{min}\{\tilde{\boldsymbol{n}} \} \to \infty$. So, we then have that for fixed $m$
\begin{equation*}
    \frac{1}{m}\sum_{s=1}^m \left(\hat{\Delta}_s - \Delta_s \right) \overset{\mathbb{P}}{\rightarrow} 0
\end{equation*}
as $\text{min}\{\tilde{\boldsymbol{n}} \} \to \infty$ if our models are correctly specified and regularity conditions (i) to (iv) hold.

Because we are able to show the result in Equation \ref{Eq: Last_hard_bit} for fixed $m$, it seems reasonable to expect that it should also hold for $m \to \infty$ so long as $m$ is growing at a rate that is appropriately bounded. We conjecture that the rate needed is similar to $m = o\left(\frac{n}{\text{log}(n)}\right)$ based on an application of results in He and Shao \cite{he2000parameters} that shows 
\begin{equation} \label{Eq: Consist_Theta} 
\begin{split}
    & ||\hat{\boldsymbol{\theta}}_n - \boldsymbol{\theta}_0|| \overset{\mathbb{P}}{\to} 0 \\
    & ||\hat{\boldsymbol{\beta}}_n - \boldsymbol{\beta}_0|| \overset{\mathbb{P}}{\to} 0
\end{split}
\end{equation}
as $n,m \to \infty$ if $m = o\left(\frac{n}{\text{log}(n)}\right)$, regularity conditions Conditions (i)-(iv) hold, and the models specified in Equations \ref{ConAppend:treatment_mod} and Equation \ref{ConAppend: studymem_model} are correct under the following additional regularity conditions:
\begin{itemize}[label={}]
    \item \textbf{(vii)}. $\sum_{i=1}^n ||\boldsymbol{V}_i||^2 = O(nm)$
    \item \textbf{(viii)}. $\sup_{\boldsymbol{\alpha},\boldsymbol{\gamma} \in H_{qm}} \sum_{i=1}^n (\boldsymbol{\alpha}^t \boldsymbol{Z}_i)^2 (\boldsymbol{\gamma}^t \boldsymbol{Z}_i)^2 = O(n)$
    \item \textbf{(ix)}. $\sup_{\boldsymbol{\alpha},\boldsymbol{\gamma} \in H_{qm}} \sum_{i=1}^n (\boldsymbol{\alpha}^t \boldsymbol{U}_{is})^2 (\boldsymbol{\gamma}^t \boldsymbol{U}_{is})^2 = O(n)$ for all $s \in \mathcal{S}$
    \item \textbf{(x)}. $\text{lim inf}_{n\to \infty} \lambda_{min}(\boldsymbol{D}_{n,s}) > 0$ for each $s \in \mathcal{S}$ where
    \begin{equation*}
    \begin{split}
        \boldsymbol{D}_{n,s} = \frac{1}{n}\sum_{i=1}^n I(S_i=s)  \frac{\text{exp}\{\boldsymbol{V}_i^t\boldsymbol{\theta}_{0,s}\}}{1+\text{exp}\{\boldsymbol{V}_i^t\boldsymbol{\theta}_{0,s}\}} \boldsymbol{V}_i^t\boldsymbol{V}_i.
    \end{split}
    \end{equation*}
    \item \textbf{(xi)}. $\text{lim inf}_{n\to \infty} \lambda_{min}(\boldsymbol{G}_{n}) > 0$ where
    \begin{equation*}
    \begin{split}
        \boldsymbol{G}_{n} = \frac{1}{n}\sum_{i=1}^n\sum_{s\in \mathcal{S}}  \frac{\text{exp}\{\boldsymbol{U}_{i,s}^t\boldsymbol{\beta}_{0}\}}{1+\text{exp}\{\boldsymbol{U}_{i,s}^t\boldsymbol{\beta}_{0}\}} \boldsymbol{U}_{i,s}^t\boldsymbol{U}_{i,s}.
    \end{split}
    \end{equation*}
\end{itemize}
where $||\boldsymbol{a}||$ is the $L_2$ norm of $\boldsymbol{a}$, $H_{qm} = \{\boldsymbol{\alpha} \in \mathbb{R}^{qm} : ||\boldsymbol{\alpha}|| =1\}$, and $\lambda_{min}$ denotes the smallest eigenvalue of a matrix. For a sketch of the application of the results from  He and Shao \cite{he2000parameters} to our setting, see Lemma 2 in Section \ref{Sec: Const_Support}.

\subsection{Supporting Results} \label{Sec: Const_Support}

\noindent \textbf{Lemma 1}: \emph{For a fixed number of studies $m$,
\begin{enumerate}[label=(\roman*)]
    \item $\frac{1}{n_s}\sum_{i: S_i =s}\hat{w}_i(a,s) \overset{\mathbb{P}}{\longrightarrow} E_{A,X}\left(w(a,s)|S=s\right)$ 
    \item $\frac{1}{n_s}\sum_{i: S_i =s}\hat{w}_i(a,s)Y_i \overset{\mathbb{P}}{\longrightarrow} E_{A,X,Y}\left(w(a,s)Y|S=s\right)$.
\end{enumerate}
for each $s=1,2,...,m$ and $a \in \mathcal{A}$ if $\hat{e}_a(\boldsymbol{V}_i,s) \overset{\mathbb{P}}{\to} e_a(\boldsymbol{X}_i,s)$ and $\frac{\hat{p}(\boldsymbol{X}_i,s)}{\hat{p}(\boldsymbol{X}_i,s)}\overset{\mathbb{P}}{\to} \frac{p(\boldsymbol{X}_i,s)}{p(\boldsymbol{X}_i,s)}$ as $\text{min}\{\tilde{\boldsymbol{n}} \} \to \infty$ where $\tilde{\boldsymbol{n}} = \{n_s\}_{s=0}^m$.}

\noindent \underline{Proof of Lemma 1}: For each $a \in \mathcal{A}$ and $s =1,2,...,m$, define 
\begin{equation*}
    \hat{\theta}_i (a,s)= \frac{e_a(\boldsymbol{X}_i,s)}{\hat{e}_a(\boldsymbol{X}_i,s)}\frac{\hat{p}(\boldsymbol{X}_i,0)}{p(\boldsymbol{X}_i,0)}\frac{p(\boldsymbol{X}_i,s)}{\hat{p}(\boldsymbol{X}_i,s)}.
\end{equation*}
Let $\hat{\theta}_{[j]}(a,s)$ is the $j$th largest $\hat{\theta}_i(a,s)$ for $i \in \{1,2,...,n:S_i=s\}$ and define 
\begin{equation*}
\begin{split}
    &   M_{n}(a,s) = \text{max}\left\{ \left|C_{[1]}(a,s) \right|, \left|C_{[n_s]}(a,s) \right| \right\} \\
    & M^*_{n}(a,s) =  \text{max}\left\{ \left|C^*_{[1]}(a,s) \right|, \left|C^*_{[n_s]}(a,s) \right| \right\} \\
\end{split}
\end{equation*}
for all $j=1,2,...,n_s$, where
\begin{equation*}
\begin{split}
    & C_{[j]}(a,s) = \frac{1}{n_s}\sum_{i: S_i=s} \hat{\theta}_{[j]}(a,s)w_i(a,s) -  E \left(w(a,s)|S=s\right) \\
    & C^*_{[j]}(a,s) = \frac{1}{n_s}\sum_{i:S_i=s} \hat{\theta}_{[j]}(a,s)w_i(a,s)Y_i -  E \left(w(a,s)Y|S=s\right). 
\end{split}
\end{equation*}
For each $a \in \mathcal{A}$ and $s =1,2,...,m$, 
\begin{equation*}
\begin{split}
   & \left|\frac{1}{n_s}\sum_{i: S_i=s} \hat{\theta}_{i}(a,s) w_i(a,s)-  E_{A,X}\left(w(a,s)
   |S=s\right) \right| \leq M_n(a,s) \\
    & \Rightarrow P\left(\left|\frac{1}{n_s}\sum_{i: S_i=s} \hat{\theta}_{i}(a,s)w_i(a,s) -  E_{A,X}\left(w(a,s)|S=s \right)\right| > \epsilon \right) \leq P(M_n(a,s) > \epsilon) \\
 \end{split}
\end{equation*}
for all $\epsilon > 0$. So, by Sub-Lemma 2.3, we have that 
\begin{equation*}
\begin{split}    
    & P\left(\left|\frac{1}{n_s}\sum_{i: S_i=s} \hat{\theta}_{i}(a,s)w_i(a,s) - E_{A,X} \left(w(a,s)|S=s \right)\right| > \epsilon \right) \to 0 \\
\end{split}
\end{equation*}
for all $\epsilon > 0$ as as $n \to \infty$ such that $n_s \to \infty$ for all $s =1,2,...,m$. This implies that
\begin{equation*}
     \frac{1}{n_s}\sum_{i: S_i=s} \hat{\theta}_{i}(l,s)w_i(l,s) \overset{\mathbb{P}}{\longrightarrow} E_{A,X}\left(w(a,s)|S=s \right)
\end{equation*}
so, by Sub-Lemma 2.1, we have 
\begin{equation*}
\begin{split}    
\frac{1}{n_s}\sum_{i: S_i=s} \hat{w}_i(l,s) \overset{\mathbb{P}}{\longrightarrow}  E_{A,X}\left(w(a,s)|S=s \right)
\end{split}
\end{equation*}
as $\text{min}\{\tilde{\boldsymbol{n}} \} \to \infty$ which proves Part (i). The proof for Part (ii) follows similarly, utilizing $C^*_{[j]}(a,s)$ and $M_n^*$ 

\qed

\noindent \textbf{Sub-Lemma 1.1} \emph{For each for each $s =1,2,...,m$ and $a\in \mathcal{A}$, 
\begin{enumerate}[label=(\roman*)]
    \item $\frac{1}{n_s}\sum_{i: S_i =s}\hat{w}_i(a,s)  = \frac{1}{n_s}\sum_{i: S_i =s} \hat{\theta}_i(a,s)w_i(a,s)$ 
    \item $\frac{1}{n_s}\sum_{i: S_i =s}\hat{w}_i(a,s) Y_i= \frac{1}{n_s}\sum_{i: S_i =s}\hat{\theta}_i(a,s)w_i(a,s)Y_i  $ 
\end{enumerate}
where 
\begin{equation*}
    \hat{\theta}_i (a,s)= \frac{e_a(\boldsymbol{X}_i,s)}{\hat{e}_a(\boldsymbol{X}_i,s)}\frac{\hat{p}(\boldsymbol{X}_i,0)}{p(\boldsymbol{X}_i,0)}\frac{p(\boldsymbol{X}_i,s)}{\hat{p}(\boldsymbol{X}_i,s)}
\end{equation*}
}

\noindent \underline{Proof of Sub-Lemma 1.1}: For each for each $s =1,2,...,m$, $a\in  \mathcal{A}$, and $i =1,2,..,n$, we can write 
\begin{equation*}
\begin{split}
    \hat{w}_i(a,s) = &  \left[\frac{I(A_i=a)}{e_a(\boldsymbol{X}_i,s)} \frac{e_a(\boldsymbol{X}_i,s)}{\hat{e}_a(\boldsymbol{X}_i,s)}\right]\left[\frac{p(\boldsymbol{X}_i,0)}{p(\boldsymbol{X}_i,s)}\frac{\hat{p}(\boldsymbol{X}_i,0)}{p(\boldsymbol{X}_i,0)}\frac{p(\boldsymbol{X}_i,s)}{\hat{p}_s(\boldsymbol{X}_i,s)}\right]  \\
    = & \frac{e_a(\boldsymbol{X}_i,s)}{\hat{e}_a(\boldsymbol{X}_i,s)}\frac{\hat{p}(\boldsymbol{X}_i,0)}{p(\boldsymbol{X}_i,0)}\frac{p(\boldsymbol{X}_i,s)}{\hat{p}(\boldsymbol{X}_i,s)} \left[\frac{I(A_i=a)}{e_a(\boldsymbol{X}_i,s)} \frac{p(\boldsymbol{X}_i,0)}{p(\boldsymbol{X}_i,s)}\right]  \\
    = & \hat{\theta}_i(a,s) w_i(a,s) \\
\end{split}
\end{equation*}
which proves (i) and (ii).

\qed

\noindent \textbf{Sub-Lemma 1.2} \emph{For a fixed number of studies $m$, 
\begin{equation*}
    \hat{\theta}_i (a,s)\overset{\mathbb{P}}{\longrightarrow} 1 
\end{equation*} 
for each $s=1,2,...,m$ and $a\in \mathcal{A}$ as $\text{min}\{\tilde{\boldsymbol{n}} \} \to \infty$.}

\noindent\underline{Proof of Sub-Lemma 1.2}: From Slutkzy's Theorem, we have that
\begin{equation*}
\begin{split}
    & \frac{e_a(\boldsymbol{X}_i,s)}{\hat{e}_a(\boldsymbol{X}_i,s)} \overset{\mathbb{P}}{\longrightarrow}  1  \\
    & \frac{\hat{p}(\boldsymbol{X}_i,0)}{p(\boldsymbol{X}_i,0)}\frac{p(\boldsymbol{X}_i,s)}{\hat{p}(\boldsymbol{X}_i,s)} = \frac{\frac{\hat{p}(\boldsymbol{X}_i,0)}{\hat{p}(\boldsymbol{X}_i,s)}}{\frac{p(\boldsymbol{X}_i,0)}{p(\boldsymbol{X}_i,s)}} \overset{\mathbb{P}}{\longrightarrow} 1
\end{split}
\end{equation*}
as $\text{min}\{\tilde{\boldsymbol{n}} \} \to \infty$. We then have 
\begin{equation*}
    \hat{\theta}_i (a,s) \overset{\mathbb{P}}{\longrightarrow} 1 
\end{equation*}
as $\text{min}\{\tilde{\boldsymbol{n}} \} \to \infty$. 

\qed

\noindent \textbf{Sub-Lemma 1.3} \emph{For fixed $m$,
\begin{enumerate}[label=(\roman*)]
    \item $\forall \epsilon > 0$, $P(M_n(a,s) > \epsilon) \to 0$ 
    \item $\forall \epsilon > 0$, $P(M^*_n(a,s) > \epsilon) \to 0$ 
\end{enumerate}
as $\text{min}\{\tilde{\boldsymbol{n}} \} \to \infty$ and $a\in \mathcal{A}$, where
\begin{equation*}
\begin{split}
    &   M_n(a,s) = \text{max}\left\{ \left|C_{[1]}(a,s) \right|, \left|C_{[n_s]}(a,s) \right| \right\} \\
    & M^*_n(a,s) =  \text{max}\left\{ \left|C^*_{[1]}(a,s) \right|, \left|C^*_{[n_s]}(a,s) \right| \right\} \\
\end{split}
\end{equation*}
and for all $j=1,2,...,n_s$
\begin{equation*}
\begin{split}
    & C_{[j]}(a,s) = \frac{1}{n_s}\sum_{i: S_i=s} \hat{\theta}_{[j]}(a,s)w_i(a,s) -  E \left(w(a,s)|S=s\right) \\
    & C^*_{[j]}(a,s) = \frac{1}{n_s}\sum_{i:S_i=s} \hat{\theta}_{[j]}(a,s)w_i(a,s)Y_i -  E \left(w(l,s)Y|S=s\right)
\end{split}
\end{equation*}
and $\hat{\theta}_{[j]}(a,s)$ is the $j$th largest $\hat{\theta}_i(a,s)$ for $i \in \{1,2,...,n:S_i=s\}$.}

\noindent\underline{Proof of Sub-Lemma 2.3}: For each $a \in \mathcal{A}$ and $s =1,2,...,m$, 
\begin{equation*}
\begin{split}
   & \frac{1}{n_s} \sum_{i: S_i=s} w_i(a,s) \overset{\mathbb{P}}{\longrightarrow} E \left(w(a,s) |S=s\right) 
\end{split}
\end{equation*}
and 
\begin{equation*}
\begin{split}
   &\frac{1}{n_s} \sum_{i: S_i=s} w_i(a,s)Y_i \overset{\mathbb{P}}{\longrightarrow} E \left(w(a,s)Y|S=s\right) 
\end{split}
\end{equation*}
as $\text{min}\{\tilde{\boldsymbol{n}} \} \to \infty$ under Condition (i) and the WWLN. By Sub-Lemma 1.2 and Slutzky's Theorem, we have for $j=1,2,...,n_s$,
\begin{equation} \label{Eq: SubL2.3_1}
\begin{split}
   &\hat{\theta}_{[j]}(a,s)\left[\frac{1}{n_s} \sum_{i: S_i=s} w_i(a,s)\right]  \overset{\mathbb{P}}{\longrightarrow} E \left(w(a,s)|S=s\right)  \\
   & \Rightarrow C_{[j]}(l,s)  \overset{\mathbb{P}}{\longrightarrow} 0  \\
   &\Rightarrow  P\left(\left|C_{[j]}(l,s)  \right| > \epsilon \right) \to 0 \text{ for all } \epsilon > 0  \\
\end{split}
\end{equation}
as $\text{min}\{\tilde{\boldsymbol{n}} \} \to \infty$. Similarly, we have 
\begin{equation}  \label{Eq: SubL2.3_2}
\begin{split}
   & P\left(\left|C_{[j]}(a,s)  \right| > \epsilon \right) \to 0 \text{ for all } \epsilon > 0  \\
\end{split}
\end{equation}
as $\text{min}\{\tilde{\boldsymbol{n}} \} \to \infty$. Since Equations (\ref{Eq: SubL2.3_1}) and (\ref{Eq: SubL2.3_2}) apply for all $j$, we have that 
\begin{equation*}
\begin{split}
   & P(M_n(a,s) > \epsilon) \to 0  \\
   & P(M^*_n(a,s) > \epsilon) \to 0 
\end{split}
\end{equation*}
as $\text{min}\{\tilde{\boldsymbol{n}} \} \to \infty$ for all $\epsilon> 0$. 

\qed

\noindent \textbf{Lemma 2}: \emph{Assume that the models specified in Equation \ref{ConAppend:treatment_mod} and Equation \ref{ConAppend: studymem_model} are correct, Conditions (i)-(iv) and (vii)-(xi) hold, and $m = o\left(\frac{n}{\text{log}(n)}\right)$, so the number of studies grows with the sample size at a bounded rate. Then,
\begin{enumerate}[label=(\roman*)]
    \item $||\hat{\boldsymbol{\theta}}_n - \boldsymbol{\theta}_0|| \overset{\mathbb{P}}{\to} 0$ 
    \item $||\hat{\boldsymbol{\beta}}_n - \boldsymbol{\beta}_0|| \overset{\mathbb{P}}{\to} 0$
\end{enumerate}
as $n,m \to \infty$ where $\hat{\boldsymbol{\theta}}_n$ is the MLE of $\boldsymbol{\theta}_0$ and $\hat{\boldsymbol{\beta}}_n$ is the MLE of $\boldsymbol{\beta}_0$.}

\noindent \underline{Proof of Lemma 2}: From Sub-Lemmas 2.1 and 2.2, we have that $||\hat{\boldsymbol{\theta}}_n - \boldsymbol{\theta}_0||^2 = O_P\left(\frac{m}{n}\right)$ and $||\hat{\boldsymbol{\beta}}_n - \boldsymbol{\beta}_0||^2 = O_P\left(\frac{m}{n}\right)$ respectively. Because $m = o\left(\frac{n}{\text{log}(n)}\right)$, we know that for all $C >0$, there exists an integer $n_0 \geq 1$ such that
\begin{equation*}
    m < c \frac{n}{\text{log}(n)}
\end{equation*}
for all $n \geq n_0$. Notice that for $n \geq 3$, $\text{log}(n) > 1$ and $\frac{n}{\text{log}(n)} < n$. So, for all $n \geq n_1 = \text{max}\{3,n_0\}$, 
\begin{equation*}
    m < C \frac{n}{\text{log}(n)} < Cn
\end{equation*}
for all $n > n_1$. Therefore, 
\begin{equation} \label{Eq: movern_op1}
\begin{split}
   & m = o(n) \\
   & \Rightarrow \frac{m}{n} = o(1) \\
   & \Rightarrow \frac{m}{n} = o_P(1).
\end{split}
\end{equation}

Then, we have that as $n,m \to \infty$ with $m = o\left(\frac{n}{\text{log}(n)}\right)$
\begin{equation} \label{Eq: NormTheta_Conv}
\begin{split}
     & ||\hat{\boldsymbol{\theta}}_n - \boldsymbol{\theta}_0||^2 = O_P\left(\frac{m}{n}\right); \emph{ by Sub-Lemma 2.1} \\
     & \Rightarrow \frac{n}{m}||\hat{\boldsymbol{\theta}}_n - \boldsymbol{\theta}||^2 = O_P(1) \\
     & \Rightarrow  ||\hat{\boldsymbol{\theta}}_n - \boldsymbol{\theta}_0||^2 = \frac{m}{n}\frac{n}{m}||\hat{\boldsymbol{\theta}}_n - \boldsymbol{\theta}_0||^2 = o_P(1)O_P(1); \emph{ by Equation \ref{Eq: movern_op1}} \\
     & \Rightarrow  ||\hat{\boldsymbol{\theta}}_n - \boldsymbol{\theta}_0||^2 =o_P(1)\\
     & \Rightarrow ||\hat{\boldsymbol{\theta}}_n - \boldsymbol{\theta}_0||^2 \overset{\mathbb{P}}{\to} 0
\end{split}
\end{equation}
and we can similarly show that as $n,m \to \infty$ with $m = o\left(\frac{n}{\text{log}(n)}\right)$
\begin{equation*}
\begin{split}
     & ||\hat{\boldsymbol{\beta}}_n - \boldsymbol{\beta}_0||^2 \overset{\mathbb{P}}{\to} 0
\end{split}
\end{equation*}
because we have that $||\hat{\boldsymbol{\beta}}_n - \boldsymbol{\beta}_0||^2 = O_P\left(\frac{m}{n}\right)$ from Sub-Lemma 2.2. 

\qed 

\noindent \textbf{Sub-Lemma 2.1}: \emph{If the models in Equation \ref{ConAppend:treatment_mod} are correctly specified, Conditions (i), (iii), (vii), (viii), and (x) hold, then 
\begin{equation*}
 ||\hat{\boldsymbol{\theta}}_n - \boldsymbol{\theta}_0||^2 = O_P\left(\frac{m}{n}\right)   
\end{equation*}
if the number of studies, $m$, grows with the sample size, $n$, with $m = o\left(\frac{n}{\text{log}(n)}\right)$ by Theorem 2.1 of He and Shao \cite{he2000parameters}, where $\hat{\boldsymbol{\theta}}_n$ is the MLE of $\boldsymbol{\theta}_0$.}

\noindent \underline{Proof of Sub-Lemma 2.1}:  Convergence results for M-estimators, such as MLEs for parameters of GLMs, when the number of parameters grows with the sample size are provided by He and Shao \cite{he2000parameters}. Because we were able to write our $m$ study-specific treatment models in Equation \ref{ConAppend:treatment_mod} as a single model with $m \times q$ parameters contained in $\boldsymbol{\theta}_0$ where $m$ grows with $n$, we only to confirm the conditions of He and Shao \cite{he2000parameters} Theorem 2.1 hold to prove Sub-Lemma 2.1. We do this by following Example 3 in He and Shao \cite{he2000parameters}. 

First, we confirm that $\hat{\theta}_n$ is a minimizer of $\sum_{i=1}^n \rho((A_i,\boldsymbol{V}_i),\boldsymbol{\theta})$ over $\boldsymbol{\theta}\in \mathbb{R}^{qm}$ for some function $\rho((A_i,\boldsymbol{V}_i),\boldsymbol{\theta})$ that is convex in $\theta$. For the model shown in Equation \ref{ConAppend:treatment_mod}, the log-likelihood is
\begin{equation*}
\begin{split}
    \text{log}(L(\boldsymbol{\theta})) = & \text{log}\left[\prod_{i=1}^n\left(\frac{\text{exp}\{\boldsymbol{Z}_i^t\boldsymbol{\theta}\}}{1+\text{exp}\{\boldsymbol{Z}_i^t\boldsymbol{\theta}\}}\right)^{I(A_i=a)}\left(\frac{1}{1+\text{exp}\{\boldsymbol{Z}_i^t\boldsymbol{\theta}\}}\right)^{1-I(A_i=a)}\right]\\
    = & \text{log}\left[\prod_{i=1}^n\frac{(\text{exp}\{\boldsymbol{Z}_i^t\boldsymbol{\theta}\})^{I(A_i=a)}}{1+\text{exp}\{\boldsymbol{Z}_i^t\boldsymbol{\theta}\}}\right]\\
    = & \sum_{i=1}^n \left[I(A_i=a)\text{log}\left(\text{exp}\{\boldsymbol{Z}_i^t\boldsymbol{\theta}_0\}\right) - \text{log}\left(1+\text{exp}\{\boldsymbol{Z}_i^t\boldsymbol{\theta}\}\right)\right]\\
    = & \sum_{i=1}^n \left[I(A_i=a)\left(\boldsymbol{Z}_i^t\boldsymbol{\theta}\right) - \text{log}\left(1+\text{exp}\{\boldsymbol{Z}_i^t\boldsymbol{\theta}\}\right)\right].\\
\end{split}
\end{equation*}
Let $\rho((A_i,\boldsymbol{V}_i),\boldsymbol{\theta}) = I(A_i=a)\left(\boldsymbol{Z}_i^t\boldsymbol{\theta}\right) - \text{log}\left(1+\text{exp}\{\boldsymbol{Z}_i^t\boldsymbol{\theta}\}\right)$ which is convex in $\boldsymbol{\theta}$ \citep{he2000parameters}. By the definition of MLEs, $\hat{\boldsymbol{\theta}}_n$ minimizes $\sum_{i=1}^n \rho((A_i,\boldsymbol{V}_i),\boldsymbol{\theta})$ over $\boldsymbol{\theta}\in \mathbb{R}^{qm}$. Letting $ \psi((A_i,\boldsymbol{V}_i),\boldsymbol{\theta}) \equiv \frac{\partial}{\partial \boldsymbol{\theta}} \rho((A_i,\boldsymbol{V}_i),\boldsymbol{\theta})$, we have 
\begin{equation} \label{Eq: const_psi}
\begin{split}
    \psi((A_i,\boldsymbol{V}_i),\boldsymbol{\theta}) = &  \frac{\partial}{\partial \boldsymbol{\theta}}\left[I(A_i=a)\left(\boldsymbol{Z}_i^t\boldsymbol{\theta}\right) - \text{log}\left(1+\text{exp}\{\boldsymbol{Z}_i^t\boldsymbol{\theta}\}\right)\right] \\
    = & \left(I(A_i=a) - \frac{\text{exp}\{\boldsymbol{Z}_i^t\boldsymbol{\theta}\}}{1+\text{exp}\{\boldsymbol{Z}_i^t\boldsymbol{\theta}\}}\right)\boldsymbol{Z}_i
\end{split}
\end{equation}

We now verify the six conditions He and Shao \cite{he2000parameters} assumed held when proving Theorem 2.1. The first condition is:
\begin{itemize}[label={}]
\item \textbf{R0}. $||\sum_{i=1}^n \psi((A_i,\boldsymbol{V}_i), \hat{\boldsymbol{\theta}}_n)|| = o_P(n^{\frac{1}{2}})$
\end{itemize}
Because $\rho$ is differentiable with respect to $\boldsymbol{\theta}$, Condition R0 is met \citep{he2000parameters}. 

The second condition is:
\begin{itemize}[label={}]
\item \textbf{R1}. $\exists C$ and $r \in (0,2]$ such that 
\begin{equation*}
\max_{i \leq n} \left\{ E_{\theta}\left[\sup_{\boldsymbol{\tau}:||\boldsymbol{\tau}-\boldsymbol{\theta}||\leq d} \left(||\eta_i(\boldsymbol{\tau},\boldsymbol{\theta})||^2 \right)\right]\right\} \leq n^Cd^r 
\end{equation*}
for $0 < d \leq 1$ where 
\begin{equation*}
\begin{split}
    \eta_i(\boldsymbol{\tau},\boldsymbol{\theta}) = &  \psi((A_i,\boldsymbol{V}_i),\boldsymbol{\tau}) - \psi((A_i,\boldsymbol{V}_i), \boldsymbol{\theta}) \\
    & - E\left[\psi((A_i,\boldsymbol{V}_i),\boldsymbol{\tau}) \right]+E\left[\psi((A_i,\boldsymbol{V}_i),\boldsymbol{\theta}) \right] \\
\end{split}
\end{equation*}
\end{itemize}

To show that Condition R1 holds, notice that 
\begin{equation} \label{Eq: normZ_normV}
\begin{split}
    ||\boldsymbol{Z}_i||^2 = \boldsymbol{Z}_i^t\boldsymbol{Z}_i = \sum_{s \in \mathcal{S}} \boldsymbol{Z}_{is}^t\boldsymbol{Z}_{is} = \sum_{s\in \mathcal{S}} I(S_i=s) \boldsymbol{V}_{i}^t\boldsymbol{V}_{i} =  \boldsymbol{V}_{i}^t\boldsymbol{V}_{i} = ||\boldsymbol{V}_i||^2
\end{split}
\end{equation}
and, following from Equation \ref{Eq: const_psi},
\begin{equation*}
\begin{split}
    \eta_i(\boldsymbol{\tau},\boldsymbol{\theta}) = & \left(\frac{\text{exp}\{\boldsymbol{Z}_i^t\boldsymbol{\theta}\}}{1+\text{exp}\{\boldsymbol{Z}_i^t\boldsymbol{\theta}\}} - \frac{\text{exp}\{\boldsymbol{Z}_i^t\boldsymbol{\tau}\}}{1+\text{exp}\{\boldsymbol{Z}_i^t\boldsymbol{\tau}\}}\right)\boldsymbol{Z}_i \\
    & +E\left[\left(\frac{\text{exp}\{\boldsymbol{Z}_i^t\boldsymbol{\tau}\}}{1+\text{exp}\{\boldsymbol{Z}_i^t\boldsymbol{\tau}\}} - \frac{\text{exp}\{\boldsymbol{Z}_i^t\boldsymbol{\theta}\}}{1+\text{exp}\{\boldsymbol{Z}_i^t\boldsymbol{\theta}\}}\right)\boldsymbol{Z}_i\right]. \\ 
\end{split}
\end{equation*}

The first derivative of $\text{expit}(x) = \frac{\text{exp}\{x\}}{1+\text{exp}\{x\}}$ is $\frac{\text{exp}\{x\}}{(1+\text{exp}\{x\})^2}$. For all $x$, $\text{exp}\{x\} > 0$ and the function $f(y) = \frac{y}{(1+y)^2}$ is bounded below by 0 and above by 0.25 (its global maximum at $y=1$) for $y>0$. Therefore, the first derivative of the expit function exists and is bounded, making the expit function a Lipschitz function. Thus, there exists a constant $C_1 >0$ such that, for all $i$,  
\begin{equation*}
\begin{split}
    \left|\frac{\text{exp}\{\boldsymbol{Z}_i^t\boldsymbol{\theta}\}}{1+\text{exp}\{\boldsymbol{Z}_i^t\boldsymbol{\theta}\}} - \frac{\text{exp}\{\boldsymbol{Z}_i^t\boldsymbol{\tau}\}}{1+\text{exp}\{\boldsymbol{Z}_i^t\boldsymbol{\tau}\}}\right| \leq & C_1 \left|\boldsymbol{Z}_i^t\boldsymbol{\theta} - \boldsymbol{Z}_i^t \boldsymbol{\tau}\right| \\
    \leq &  C_1 ||\boldsymbol{Z}_i|| ||\boldsymbol{\theta}-\boldsymbol{\tau}|| ; \emph{ Cauchy-Schwarz}  \\
    = & C_1||\boldsymbol{V}_i|| ||\boldsymbol{\theta}-\boldsymbol{\tau}||; \emph{ Equation \ref{Eq: normZ_normV}}.
\end{split}
\end{equation*}
So, for all $i$, 
\begin{equation} \label{Eq: norm_diff}
\begin{split}
  \left|\left|\left(\frac{\text{exp}\{\boldsymbol{Z}_i^t\boldsymbol{\theta}\}}{1+\text{exp}\{\boldsymbol{Z}_i^t\boldsymbol{\theta}\}} - \frac{\text{exp}\{\boldsymbol{Z}_i^t\boldsymbol{\tau}\}}{1+\text{exp}\{\boldsymbol{Z}_i^t\boldsymbol{\tau}\}}\right)\boldsymbol{Z}_i\right|\right| = &  \left|\frac{\text{exp}\{\boldsymbol{Z}_i^t\boldsymbol{\theta}\}}{1+\text{exp}\{\boldsymbol{Z}_i^t\boldsymbol{\theta}\}} - \frac{\text{exp}\{\boldsymbol{Z}_i^t\boldsymbol{\tau}\}}{1+\text{exp}\{\boldsymbol{Z}_i^t\boldsymbol{\tau}\}}\right| ||\boldsymbol{Z}_i|| \\
  \leq & C_1 ||\boldsymbol{V}_i||^2 ||\boldsymbol{\theta}-\boldsymbol{\tau}|| 
\end{split}
\end{equation}
Let $M_{i}= \text{max}\left\{||\boldsymbol{V}_i||^2, E\left(||\boldsymbol{V}_i||^2\right)\right\}= \text{max}\left\{||\boldsymbol{Z}_i||^2, E\left(||\boldsymbol{Z}_i||^2\right)\right\}$ and notice that $M_i \geq 0$. Then, from Equation \ref{Eq: norm_diff} and Jensen's inequality we have
\begin{equation*}
\begin{split}
     ||\eta_i(\boldsymbol{\tau},\boldsymbol{\theta})||\leq & C_1||\boldsymbol{V}_i||^2||\boldsymbol{\tau}-\boldsymbol{\theta}|| +E\left( C_1||\boldsymbol{V}_i||^2||\boldsymbol{\tau}-\boldsymbol{\theta}||\right) \\
     = & C_1\left[||\boldsymbol{V}_i||^2 + E\left(||\boldsymbol{V}_i||^2\right)\right]||\boldsymbol{\tau}-\boldsymbol{\theta}|| \\
\end{split}
\end{equation*}
and then that 
\begin{equation} \label{ConsistencAppend: eta_log}
\begin{split}
     ||\eta_i(\boldsymbol{\tau},\boldsymbol{\theta})|| \leq & 2C_1M_i ||\boldsymbol{\tau}-\boldsymbol{\theta}||
\end{split}
\end{equation}

Notice that $\left|\left|\frac{\boldsymbol{Z}^t_i}{||\boldsymbol{Z}_i||} \right|\right| = 1$ so by Assumption (viii), letting $\boldsymbol{\alpha} = \boldsymbol{\gamma} = \frac{\boldsymbol{Z}^t_i}{||\boldsymbol{Z}_i||}$, 
\begin{equation} \label{Eq: Eq_B11}
\begin{split}
    & \sum_{i=1}^n \left|\frac{\boldsymbol{Z}^t_i}{||\boldsymbol{Z}_i||}\boldsymbol{Z}_i \right|^2 \left|\frac{\boldsymbol{Z}^t_i}{||\boldsymbol{Z}_i||}\boldsymbol{Z}_i \right|^2 = \sum_{i=1}^n  ||\boldsymbol{Z}_i||^4 = O(n) \\
\end{split}
\end{equation}
By Jensen' inequality and the linearity of expectations, we then have that
\begin{equation} \label{Eq: Eq_B12}
\begin{split}
   & \sum_{i=1}^n E\left(||\boldsymbol{Z}_i||^2\right)^2 \leq \sum_{i=1}^n E\left(||\boldsymbol{Z}_i||^4\right) = E\left(\sum_{i=1}^n ||\boldsymbol{Z}_i||^4\right) = O(n). \\
\end{split}
\end{equation}
Because $M_i \geq 0$, $M^2_{i}= \text{max}\left\{||\boldsymbol{Z}_i||^4, E\left(||\boldsymbol{Z}_i||^2\right)^2\right\}$, so by Equations \ref{Eq: Eq_B11} and \ref{Eq: Eq_B12} we have that 
\begin{equation} \label{ConsistencyAppend: sumM2_On}
\sum_{i=1}^n M_{i}^2 = O(n)   
\end{equation}
and then we have that 
\begin{equation} \label{ConsistencyAppend: Mi_O}
\begin{split}
    & M_i^2 \leq \sum_{i=1}^n M_{i}^2 = O(n)  \\
    & \Rightarrow M_i = O\left(n^{\frac{1}{2}}\right)
\end{split}
\end{equation}

From Example 1 in \cite{he2000parameters}, we then have that Condition R1 holds because 
\begin{equation} 
\begin{split}
     ||\eta_i(\boldsymbol{\tau},\boldsymbol{\theta})|| \leq & 2C_1M_i ||\boldsymbol{\tau}-\boldsymbol{\theta}||
\end{split}
\end{equation}
where $M_i = O\left(n^{r_1}\right)$ for some $r_1 >0$. 

The third condition is:
\begin{itemize}[label={}]
\item \textbf{R2}. $\sum_{i=1}^n E\left(||\psi((A_i,\boldsymbol{V}_i),\boldsymbol{\theta}_0)||^2\right) = O(nmq)$
\end{itemize}
Condition R2 holds because 
\begin{equation*}
\begin{split}
    \sum_{i=1}^n||\psi((A_i,\boldsymbol{V}_i).\boldsymbol{\theta}_0)||^2 = & \sum_{i=1}^n\left(I(A_i=a) - \frac{\text{exp}\{\boldsymbol{Z}_i^t\boldsymbol{\theta}_0\}}{1+\text{exp}\{\boldsymbol{Z}_i^t\boldsymbol{\theta}_0\}}\right)^2||\boldsymbol{Z}_i||^2; \emph{ Equation \ref{Eq: const_psi}}\\
    \leq & \sum_{i=1}^n||\boldsymbol{Z}_i||^2 \\
    = & \sum_{i=1}^n||\boldsymbol{V}_i||^2 ; \emph{ Equation \ref{Eq: normZ_normV}} \\
\end{split}
\end{equation*}
because $-1 < I(A_i=a) - \frac{\text{exp}\{\boldsymbol{Z}_i^t\boldsymbol{\theta}_0\}}{1+\text{exp}\{\boldsymbol{Z}_i^t\boldsymbol{\theta}_0\}} < 1$. So, under Assumption (vii) we have 
\begin{equation*}
\begin{split}
    \sum_{i=1}^n E\left(||\psi((A_i,\boldsymbol{X}_i),\boldsymbol{\theta}_0)||^2\right) \leq  E\left(\sum_{i=1}^n||\boldsymbol{V}_i||^2\right) = O(nm)
\end{split}
\end{equation*}
and since $q$ is constant, $O(nm) = O(nmq)$. 

The fourth condition is:
\begin{itemize}[label={}]
\item \textbf{R3}. There exists a sequence of $(mq) \times (mq)$ matrices, $\boldsymbol{D}_n$ with $$\text{lim inf}_{n \to \infty} \lambda_{min}(\boldsymbol{D}_n)> 0$$ such that for any $B >0$ and uniformly in $\boldsymbol{\alpha} \in H_{m} = \{\boldsymbol{\alpha} \in \mathbb{R}^{m} : ||\boldsymbol{\alpha}|| =1\}$
\begin{equation*}
    \sup_{||\boldsymbol{\theta}-\boldsymbol{\theta}_0|| \leq B\left(\frac{qm}{n}\right)^{\frac{1}{2}}}\left|\boldsymbol{\alpha}^t\sum_{i=1}^n E_{\boldsymbol{\theta}_0}\left[\psi((A_i,\boldsymbol{V}_i),\boldsymbol{\theta}) - \psi((A_i,\boldsymbol{V}_i), \boldsymbol{\theta}_0)\right] - n\boldsymbol{\alpha}^t \boldsymbol{D}_n(\boldsymbol{\theta}-\boldsymbol{\theta}_0)\right|
\end{equation*}
is $ o(n^{\frac{1}{2}})$. 
\end{itemize}
Let $\boldsymbol{D}_n = \frac{1}{n} \sum_{i=1}^n \frac{\text{exp}\{\boldsymbol{Z}_i^t\boldsymbol{\theta}_0\}}{1+\text{exp}\{\boldsymbol{Z}_i^t\boldsymbol{\theta}_0\}}\boldsymbol{Z}_i\boldsymbol{Z}_i^t$. Because 
\begin{equation*}
\begin{split}
  \boldsymbol{Z}_i\boldsymbol{Z}_i^t = \text{blockdiag}\left(I(S_i=1)\boldsymbol{V}_i\boldsymbol{V}_i^t,....,I(S_i=m)\boldsymbol{V}_i\boldsymbol{V}_i^t\right)
\end{split}
\end{equation*}
and $\boldsymbol{Z}_i^t\boldsymbol{\theta}_0 = \boldsymbol{V}_i^t\boldsymbol{\theta}_{0,S_i}$, we can write 
\begin{equation*}
\begin{split}
    \boldsymbol{D}_n =\text{block}\left(\boldsymbol{D}_{n,1},...,\boldsymbol{D}_{n,m}\right)
\end{split}
\end{equation*}
where 
\begin{equation*}
\begin{split}
    \boldsymbol{D}_{n,s} = \frac{1}{n}\sum_{i=1}^n I(S_i=s)  \frac{\text{exp}\{\boldsymbol{V}_i^t\boldsymbol{\theta}_{0,s}\}}{1+\text{exp}\{\boldsymbol{V}_i^t\boldsymbol{\theta}_{0,s}\}} \boldsymbol{V}_i^t\boldsymbol{V}_i.
\end{split}
\end{equation*}
The eigenvalues of a block diagonal matrix are the eigenvalues of each block, so by Assumption (x), we have that $\text{lim inf}_{n\to \infty} \lambda_{min}(\boldsymbol{D}_n) > 0$. Then, from Example 3 in \cite{he2000parameters} we have that Condition R3 holds under Assumptions (vii) and (viii). 

The fifth condition is:
\begin{itemize}[label={}]
\item \textbf{R4}. For any $\boldsymbol{\theta} \in \mathbb{R}^{qm}$, $\boldsymbol{\alpha} \in H_{qm}$, and $B >0$,
\begin{equation*}
    \sup_{\boldsymbol{\tau}: ||\boldsymbol{\tau} - \boldsymbol{\theta}|| \leq B\left(\frac{mq}{n}\right)^{\frac{1}{2}}} \left\{\sum_{i=1}^n E_{\boldsymbol{\theta}}\left[|\boldsymbol{\alpha}^t\eta_i(\boldsymbol{\tau},\boldsymbol{\theta})|^2\right]\right\} = O(m)
\end{equation*}
\end{itemize}
To see that Condition R4 holds, first notice, by Cauchy–Schwarz and Equation \ref{ConsistencAppend: eta_log} we have that there exists some constant $C >0$ such that 
\begin{equation*}
\begin{split}
    \sum_{i=1}^n(\boldsymbol{\alpha}^t\eta_i(\boldsymbol{\tau},\boldsymbol{\theta}))^2 \leq & \sum_{i=1}^n||\boldsymbol{\alpha}||^2 ||\eta_i(\boldsymbol{\tau},\boldsymbol{\theta})||^2 = \sum_{i=1}^n||\eta_i(\boldsymbol{\tau},\boldsymbol{\theta})||^2 \\
    \leq & C||\boldsymbol{\tau}-\boldsymbol{\theta}||^2\sum_{i=1}^nM_{i}^2 \\
\end{split}
\end{equation*}
because $\boldsymbol{\alpha} \in H_{qm}$. So, there exists constants $C,C^* >0$ and $n_0$ such that for all $n \geq n_0$
\begin{equation*}
\begin{split}
    \sup_{\boldsymbol{\tau}: ||\boldsymbol{\tau} - \boldsymbol{\theta}|| \leq B\left(\frac{mq}{n}\right)^{\frac{1}{2}}} & \Big\{\sum_{i=1}^n E_{\boldsymbol{\theta}}\big[|\boldsymbol{\alpha}^t\eta_i(\boldsymbol{\tau},\boldsymbol{\theta})|^2\big]\Big\}\\
    \leq &\sup_{\boldsymbol{\tau}: ||\boldsymbol{\tau} - \boldsymbol{\theta}|| \leq B\left(\frac{mq}{n}\right)^{\frac{1}{2}}} \left\{C||\boldsymbol{\tau}-\boldsymbol{\theta}||^2 E_{\boldsymbol{\theta}}\left[  \sum_{i=1}^nM_{i}^2\right]\right\}; \emph{ by Equation \ref{ConsistencAppend: eta_log}} \\ 
    \leq & CB^2q\left(\frac{m}{n}\right) E_{\boldsymbol{\theta}}\left[  \sum_{i=1}^nM_{i}^2\right] \\
     \leq & C^*\left(\frac{m}{n}\right)n ; \emph {by Equation \ref{ConsistencyAppend: sumM2_On} and because } q \emph{ is a constant}\\
     = & C^*m\\
\end{split}
\end{equation*}
which proves Condition R4. 

The sixth, and final, condition is:
\begin{itemize}[label={}]
\item \textbf{R5}. For any $\boldsymbol{\theta} \in \mathbb{R}^{mq}$ and $B >0$,
\begin{equation*}
    \sup_{\boldsymbol{\alpha} \in H_{qm}}\left\{\sup_{\boldsymbol{\tau}: ||\boldsymbol{\tau} - \boldsymbol{\theta}|| \leq B\left(\frac{qm}{n}\right)^{\frac{1}{2}}}\sum_{i=1}^n\left[\boldsymbol{\alpha}^t \eta_i(\boldsymbol{\tau},\boldsymbol{\theta}) \right]^2\right\} = O_P(m)
\end{equation*}
\end{itemize}
Using the same logic as in the proof that Condition R4 holds, we have that there exists a constant $C^*>0$ and an $n_0$ such that for all $n > n_0$
\begin{equation*}
 \begin{split}
      \sup_{\boldsymbol{\alpha} \in H_{qm}}\left\{\sup_{\boldsymbol{\tau}: ||\boldsymbol{\tau} - \boldsymbol{\theta}|| \leq B\left(\frac{mq}{n}\right)^{\frac{1}{2}}}\sum_{i=1}^n\left[\boldsymbol{\alpha}^t \eta_i(\boldsymbol{\tau},\boldsymbol{\theta}) \right]^2\right\} \leq & CB^2\left(\frac{mq}{n}\right) \sum_{i=1}^nM_{i}^2\\
     \leq & C^*\left(\frac{m}{n}\right) n = C^*m \\
\end{split}   
\end{equation*}
which proves Condition R5 holds because, for all $n > n_0$,
\begin{equation*}
    P\left(\frac{1}{m} \sup_{\boldsymbol{\alpha} \in H_{qm}}\left\{\sup_{\boldsymbol{\tau}: ||\boldsymbol{\tau} - \boldsymbol{\theta}|| \leq B\left(\frac{mq}{n}\right)^{\frac{1}{2}}}\sum_{i=1}^n\left[\boldsymbol{\alpha}^t \eta_i(\boldsymbol{\tau},\boldsymbol{\theta}) \right]^2\right\} > C^*\right) = 0. 
\end{equation*}
\qed

\noindent \textbf{Sub-Lemma 2.2}: \emph{If the model in Equation \ref{ConAppend: studymem_model} is correctly specified, Conditions (i)-(iv), (vii), (ix), and (xi) hold, 
\begin{equation*}
 ||\hat{\boldsymbol{\beta}}_n - \boldsymbol{\beta}_0||^2 = O_P\left(\frac{m}{n}\right)   
\end{equation*}
if the number of studies, $m$, grows with the sample size, $n$, with $m = o\left(\frac{n}{\text{log}(n)}\right)$ by Theorem 2.1 of \cite{he2000parameters}, where $\hat{\beta}_n$ is the MLE of $\boldsymbol{\beta}_0$.}

\noindent \underline{Proof of Sub-Lemma 2.2}:  Our study membership model specified in Equation \ref{ConAppend: studymem_model} has $m \times q$ parameters parameters contained in $\boldsymbol{\beta}_0$ where $m$ grows with $n$. We only to confirm the conditions of He and Shao \cite{he2000parameters} Theorem 2.1 hold to prove Sub-Lemma 2.2 and therefore this proof follows the proof of Sub-Lemma 2.1. 

First, we confirm that $\hat{\beta}_n$ is a minimizer of $\sum_{i=1}^n \rho((S_i,\boldsymbol{V}_i),\boldsymbol{\beta})$ over $\boldsymbol{\beta} \in \mathbb{R}^{qm}$ for some function $\rho((S_i,\boldsymbol{V}_i),\boldsymbol{\beta})$ that is convex in $\beta$. For the model shown in Equation \ref{ConAppend: studymem_model}, the log-likelihood is
\begin{equation*}
\begin{split}
    \text{log}(L(\boldsymbol{\beta})) = & \text{log}\left[\prod_{i=1}^n\prod_{s\in \{0,\mathcal{S}\}} P(S_i=s|\boldsymbol{X}_i)^{I(S_i=s)}\right] \\
    = & \text{log}\left[\prod_{i=1}^n \prod_{s\in \mathcal{S}}\left(\frac{1}{1+ \sum_{k=1}^m\text{exp}\{\boldsymbol{U}^t_{ik}\boldsymbol{\beta}\}}\right)^{\sum_{s\in \{0,\mathcal{S}\}} I(S_i=s)} \text{exp}\{\boldsymbol{U}^t_{is}\boldsymbol{\beta}\}^{I(S_i=s)}\right] \\
    = & \sum_{i=1}^n \sum_{s \in \mathcal{S}}\left[I(S_i=s)\boldsymbol{U}^t_{is}\boldsymbol{\beta}-\text{log}\left(1+ \sum_{k \in \mathcal{S}}\text{exp}\{\boldsymbol{U}^t_{ik}\boldsymbol{\beta}\}\right)\right] \\
     = & \sum_{i=1}^n \left\{\left[\sum_{s \in \mathcal{S}}I(S_i=s)\boldsymbol{U}^t_{is}\right]\boldsymbol{\beta}-\text{log}\left(1+ \sum_{k \in \mathcal{S}}\text{exp}\{\boldsymbol{U}^t_{ik}\boldsymbol{\beta}\}\right)\right\}\\
\end{split}
\end{equation*}
because $\sum_{s\in \{0,\mathcal{S}\}} I(S_i=s) = 1$. Let 
\begin{equation*} 
\begin{split}
    \rho((S_i,\boldsymbol{V}_i),\boldsymbol{\theta}) = &\left[\sum_{s\in \mathcal{S}}I(S_i=s)\boldsymbol{U}^t_{is}\right]\boldsymbol{\beta}-\text{log}\left(1+ \sum_{s\in \mathcal{S}}\text{exp}\{\boldsymbol{U}^t_{is}\boldsymbol{\beta}\}\right).
\end{split}
\end{equation*}
which is convex in $\boldsymbol{\beta}$ \cite{he2000parameters}. By the definition of MLEs, $\hat{\boldsymbol{\beta}}_n$ minimizes $\sum_{i=1}^n \rho((S_i,\boldsymbol{V}_i),\boldsymbol{\beta}) $ over $\boldsymbol{\beta} \in \mathbb{R}^{qm}$. Letting $  \psi((S_i,\boldsymbol{V}_i),\boldsymbol{\beta})  = \frac{\partial}{\partial \boldsymbol{\beta}}\rho((S_i,\boldsymbol{V}_i),\boldsymbol{\beta})$, we have 
\begin{equation} \label{Eq: consistency_psi2}
\begin{split}
   \psi((S_i,\boldsymbol{V}_i),\boldsymbol{\beta})  = & \sum_{s\in \mathcal{S}}I(S_i=s)\boldsymbol{U}_{is} - \sum_{s\in \mathcal{S}}\frac{\text{exp}\{\boldsymbol{U}^t_{is}\boldsymbol{\beta}\}}{1+ \sum_{k\in \mathcal{S}}\text{exp}\{\boldsymbol{U}^t_{ik}\boldsymbol{\beta}\}}\boldsymbol{U}_{is} \\
   = & \sum_{s\in \mathcal{S}} \left[I(S_i=s) - \frac{\text{exp}\{\boldsymbol{U}^t_{is}\boldsymbol{\beta}\}}{1+ \sum_{k\in \mathcal{S}}\text{exp}\{\boldsymbol{U}^t_{ik}\boldsymbol{\beta}\}}\right]\boldsymbol{U}_{is}. \\
\end{split}
\end{equation}

We now verify the six conditions He and Shao \cite{he2000parameters} assumed held when proving Theorem 2.1. The first condition is:
\begin{itemize}[label={}]
\item \textbf{R0}. $||\sum_{i=1}^n \psi((S_i,\boldsymbol{V}_i), \hat{\boldsymbol{\beta}}_n)|| = o_P(n^{\frac{1}{2}})$
\end{itemize}
Because $\rho$ is differentiable with respect to $\boldsymbol{\beta}$, Condition R0 is met \cite{he2000parameters}. 

The second condition is 
\begin{itemize}[label={}]
\item \textbf{R1}. $\exists C$ and $r \in (0,2]$ such that 
\begin{equation*}
\max_{i \leq n} \left\{ E_{\theta}\left[\sup_{\boldsymbol{\tau}:||\boldsymbol{\tau}-\boldsymbol{\theta}||\leq d} \left(||\eta_i(\boldsymbol{\tau},\boldsymbol{\beta})||^2 \right)\right]\right\} \leq n^Cd^r    
\end{equation*}
for $0 < d \leq 1$ where 
\begin{equation*}
\begin{split}
    \eta_i(\boldsymbol{\tau},\boldsymbol{\theta}) = &  \psi((S_i,\boldsymbol{V}_i),\boldsymbol{\tau}) - \psi((S_i,\boldsymbol{V}_i), \boldsymbol{\theta}) \\
    & - E\left[\psi((S_i,\boldsymbol{V}_i),\boldsymbol{\tau}) \right]+E\left[\psi((S_i,\boldsymbol{V}_i),\boldsymbol{\theta}) \right] \\
\end{split}
\end{equation*}
\end{itemize}

To show that Condition R1 holds, recall that for each $i=1,2,...,n$ and $s \in \mathcal{S}$, $\boldsymbol{U}_{is} = e_s \otimes \boldsymbol{V}_i \in \mathbb{R}^{mq}$ where $e_s$ is the $s$th column of the $m \times m$ identity matrix, so $||\boldsymbol{U}_{i,s}|| = ||\boldsymbol{V}_i||$ and $\boldsymbol{U}_{is}^t \boldsymbol{\beta} = \boldsymbol{V}_i^t \boldsymbol{\beta}_s$ for each $i=1,2,...,n$ and $s \in \mathcal{S}$. Therefore,
\begin{equation} \label{Eq: SL1.2_1}
\begin{split}
    \Big|\Big|\sum_{s\in \mathcal{S}} & \left(\frac{\text{exp}\{\boldsymbol{U}^t_{is}\boldsymbol{\beta}\}}{1+ \sum_{k=1}^m\text{exp}\{\boldsymbol{U}^t_{ik}\boldsymbol{\beta}\}}- \frac{\text{exp}\{\boldsymbol{U}^t_{is}\boldsymbol{\tau}\}}{1+ \sum_{k=1}^m\text{exp}\{\boldsymbol{U}^t_{ik}\boldsymbol{\tau}\}}\right)\boldsymbol{U}_{is}\Big|\Big| \\
    \leq & \sum_{s\in \mathcal{S}}\left|\left|\left(\frac{\text{exp}\{\boldsymbol{U}^t_{is}\boldsymbol{\beta}\}}{1+ \sum_{k=1}^m\text{exp}\{\boldsymbol{U}^t_{ik}\boldsymbol{\beta}\}}- \frac{\text{exp}\{\boldsymbol{U}^t_{is}\boldsymbol{\tau}\}}{1+ \sum_{k=1}^m\text{exp}\{\boldsymbol{U}^t_{ik}\boldsymbol{\tau}\}}\right)\boldsymbol{U}_{is}\right|\right| \\
    = & \sum_{s\in \mathcal{S}}\left|\left(\frac{\text{exp}\{\boldsymbol{U}^t_{is}\boldsymbol{\beta}\}}{1+ \sum_{k=1}^m\text{exp}\{\boldsymbol{U}^t_{ik}\boldsymbol{\beta}\}}- \frac{\text{exp}\{\boldsymbol{U}^t_{is}\boldsymbol{\tau}\}}{1+ \sum_{k=1}^m\text{exp}\{\boldsymbol{U}^t_{ik}\boldsymbol{\tau}\}}\right)\right|\left|\left|\boldsymbol{U}_{is}\right|\right| \\
    = & \left|\left|\boldsymbol{V}_{i}\right|\right|\sum_{s\in \mathcal{S}}\left|\left(\frac{\text{exp}\{\boldsymbol{V}^t_{i}\boldsymbol{\beta}_s\}}{1+ \sum_{k=1}^m\text{exp}\{\boldsymbol{V}^t_{i}\boldsymbol{\beta}_k\}}- \frac{\text{exp}\{\boldsymbol{V}^t_{i}\boldsymbol{\tau}_s\}}{1+ \sum_{k=1}^m\text{exp}\{\boldsymbol{V}^t_{i}\boldsymbol{\tau}_k\}}\right)\right| \\
\end{split}
\end{equation}

Consider the following function of $\boldsymbol{x} = (x_1,...,x_{m})^t$
\begin{equation*}
 \phi_s(\boldsymbol{x}) = \frac{\text{exp}\{x_s\}}{1+\sum_{k \in {S}}\text{exp}\{x_k\}}
\end{equation*}
The derivative of $ \phi_s(\boldsymbol{x})$ with respect to $x_s$ is 
\begin{equation*}
 \frac{\partial}{\partial x_s}\phi_s(\boldsymbol{x}) =  \frac{(1+\sum_{k \in \mathcal{S},k\ne s}\text{exp}\{x_k\})\text{exp}\{x_s\}}{\left(1+\sum_{k \in {S}}\text{exp}\{x_k\}\right)^2}
\end{equation*}
and the derivative with respect to $x_t$ $t \ne s$ is 
\begin{equation*}
\begin{split}
    \frac{\partial}{\partial x_t}\phi_s(\boldsymbol{x}) = &  \text{exp}\{x_s\} \frac{\partial}{\partial x_t}\frac{1}{1+\sum_{k \in {S},k\ne t}\text{exp}\{x_k\} + \text{exp}\{x_t\}} \\
    = & \frac{\text{exp}\{x_s\} \text{exp}\{x_t\}}{\left(1+\sum_{k \in {S}} \text{exp}\{x_k\}\right)^2}
\end{split}
\end{equation*}
All of these partial derivatives exists everywhere and are bounded below and above because $\text{exp}(x) >0$ for all $x$. So $\phi_s(\boldsymbol{x})$ is a Lipschitz function and there exists a constant $C >0$ such that 
\begin{equation*}
    |\phi(\boldsymbol{x}) - \phi(\boldsymbol{y})| \leq C ||\boldsymbol{x}-\boldsymbol{y}||
\end{equation*}
This shows that for all $s \in \mathcal{S}$, there exists a constant $C_s >0$ such that
\begin{equation*}
\begin{split}
    \Bigg|\Bigg(\frac{\text{exp}\{\boldsymbol{V}^t_{i}\boldsymbol{\beta}_s\}}{1+ \sum_{k\in \mathcal{S}}\text{exp}\{\boldsymbol{V}^t_{i}\boldsymbol{\beta}_s\}}- &\frac{\text{exp}\{\boldsymbol{V}^t_{i}\boldsymbol{\tau}_s\}}{1+ \sum_{k\in \mathcal{S}}\text{exp}\{\boldsymbol{V}^t_{i}\boldsymbol{\tau}_s\}}\Bigg)\Bigg| \\
    \leq & C_s \left|\left|\begin{pmatrix} \boldsymbol{V}^t_{i}(\boldsymbol{\beta}_1-\boldsymbol{\tau}_1) \\ \vdots \\ \boldsymbol{V}^t_{i}(\boldsymbol{\beta}_{m}-\boldsymbol{\tau}_{m})  \end{pmatrix} \right|\right| \\
    = & C_s\left( \sum_{s\in \mathcal{S}} (\boldsymbol{V}_{i}^t(\boldsymbol{\beta}_s-\boldsymbol{\tau}_s) )^2\right)^{\frac{1}{2}} \\
    \leq & C_s\left( ||\boldsymbol{V}_{i}||^2 \sum_{s\in \mathcal{S}} ||\boldsymbol{\beta}_s-\boldsymbol{\tau}_s|| ^2\right)^{\frac{1}{2}} \\
     = & C_s\left( ||\boldsymbol{V}_{i}||^2 ||\boldsymbol{\beta}-\boldsymbol{\tau}|| ^2\right)^{\frac{1}{2}}  \\
     = & C_s ||\boldsymbol{V}_{i}|| ||\boldsymbol{\beta}-\boldsymbol{\tau}||
\end{split}
\end{equation*}
and that 
\begin{equation} \label{Eq: SL1.2_2}
\begin{split}
    \sum_{s\in \mathcal{S}}\Bigg|\Bigg(\frac{\text{exp}\{\boldsymbol{V}^t_{i}\boldsymbol{\beta}_s\}}{1+ \sum_{k\in \mathcal{S}}\text{exp}\{\boldsymbol{V}^t_{i}\boldsymbol{\beta}_s\}}-& \frac{\text{exp}\{\boldsymbol{V}^t_{i}\boldsymbol{\tau}_s\}}{1+ \sum_{k\in \mathcal{S}}\text{exp}\{\boldsymbol{V}^t_{i}\boldsymbol{\tau}_s\}}\Bigg)\Bigg| \\
    \leq &  \left(\sum_{s\in \mathcal{S}} C_s\right)||\boldsymbol{V}_{i}|| ||\boldsymbol{\beta}-\boldsymbol{\tau}||  \\
\end{split}
\end{equation}

Let $M_{i}= \text{max}\left\{||\boldsymbol{V}_i||^2, E\left(||\boldsymbol{V}_i||^2\right)\right\}$ and notice that $M_i \geq 0$. By Equations \ref{Eq: SL1.2_1} and \ref{Eq: SL1.2_2}, we have that there exists a constant $C_1 > 0$  
\begin{equation*}
\begin{split}
    \Big|\Big|\sum_{s\in \mathcal{S}} & \left(\frac{\text{exp}\{\boldsymbol{U}^t_{is}\boldsymbol{\beta}\}}{1+ \sum_{k\in \mathcal{S}}\text{exp}\{\boldsymbol{U}^t_{ik}\boldsymbol{\beta}\}}- \frac{\text{exp}\{\boldsymbol{U}^t_{is}\boldsymbol{\tau}\}}{1+ \sum_{k\in \mathcal{S}}\text{exp}\{\boldsymbol{U}^t_{ik}\boldsymbol{\tau}\}}\right)\boldsymbol{U}_{is}\Big|\Big| \\
    & \leq  C_1||\boldsymbol{V}_{i}||^2 ||\boldsymbol{\beta}-\boldsymbol{\tau}|| \\
\end{split}
\end{equation*}
and by Equation \ref{Eq: consistency_psi2}, we have that
\begin{equation*}
\begin{split}
    \eta_i(\boldsymbol{\tau},\boldsymbol{\beta}) = &  \sum_{s\in \mathcal{S}} \left(\frac{\text{exp}\{\boldsymbol{U}^t_{is}\boldsymbol{\beta}\}}{1+ \sum_{k\in \mathcal{S}}\text{exp}\{\boldsymbol{U}^t_{ik}\boldsymbol{\beta}\}}- \frac{\text{exp}\{\boldsymbol{U}^t_{is}\boldsymbol{\tau}\}}{1+ \sum_{k\in \mathcal{S}}\text{exp}\{\boldsymbol{U}^t_{ik}\boldsymbol{\tau}\}}\right)\boldsymbol{U}_{is} \\
    & + E\left[\sum_{s\in \mathcal{S}} \left(\frac{\text{exp}\{\boldsymbol{U}^t_{is}\boldsymbol{\tau}\}}{1+ \sum_{k\in \mathcal{S}}\text{exp}\{\boldsymbol{U}^t_{ik}\boldsymbol{\tau}\}}- \frac{\text{exp}\{\boldsymbol{U}^t_{is}\boldsymbol{\beta}\}}{1+ \sum_{k\in \mathcal{S}}\text{exp}\{\boldsymbol{U}^t_{ik}\boldsymbol{\beta}\}}\right)\boldsymbol{U}_{is} \right]\\
\end{split}
\end{equation*}
So, by Jensen's inequality we have that 
\begin{equation} \label{Eq: MN_Bount_Eta}
\begin{split}
     ||\eta_i(\boldsymbol{\tau},\boldsymbol{\beta})||\leq & C_1||\boldsymbol{V}_i||^2||\boldsymbol{\tau}-\boldsymbol{\beta}|| +E\left( C_1||\boldsymbol{V}_i||^2||\boldsymbol{\tau}-\boldsymbol{\beta}||\right) \\
     = & C_1\left[||\boldsymbol{V}_i||^2 + E\left(||\boldsymbol{V}_i||^2||\right)\right]||\boldsymbol{\tau}-\boldsymbol{\beta}|| \\
     \leq & 2C_1M_i ||\boldsymbol{\tau}-\boldsymbol{\beta}||.
\end{split}
\end{equation}
In Equation \ref{ConsistencyAppend: Mi_O} of the proof for Sub-Lemma 2.1, we showed that $M_i = O\left(n^{\frac{1}{2}}\right)$. Then, by Example 1 in \cite{he2000parameters}, we have that Condition R1 holds because $M_i =  O\left(n^{r_1}\right)$ for some $r_1 >0$. 

The third condition is 
\begin{itemize}[label={}]
\item \textbf{R2}. $\sum_{i=1}^n E\left(||\psi((S_i,\boldsymbol{V}_i),\boldsymbol{\beta}_0)||^2\right) = O(nmq)$
\end{itemize}
Notice that for all $i=1,2,...,n$, 
\begin{equation}\label{Eq: L1.2_R2_1}
\begin{split}
    || \psi((S_i,\boldsymbol{V}_i),\boldsymbol{\beta}_0)|| \leq & \sum_{s\in \mathcal{S}} \left|I(S_i=s) - \frac{\text{exp}\{\boldsymbol{U}^t_{is}\boldsymbol{\beta}_0\}}{1+ \sum_{k\in \mathcal{S}}\text{exp}\{\boldsymbol{U}^t_{ik}\boldsymbol{\beta}_0\}}\right|||\boldsymbol{U}_{i,s}|| \\
    = &||\boldsymbol{V}_{i}|| \sum_{s\in \mathcal{S}} \left|I(S_i=s) - \frac{\text{exp}\{\boldsymbol{U}^t_{is}\boldsymbol{\beta}_0\}}{1+ \sum_{k\in \mathcal{S}}\text{exp}\{\boldsymbol{U}^t_{ik}\boldsymbol{\beta}_0\}}\right| \\
\end{split}
\end{equation}
because $||\boldsymbol{U}_{i,s}|| = ||\boldsymbol{V}_i||$ for all $s \in \mathcal{S}$. Recall that we have also assumed that 
\begin{equation*} 
\begin{split}
    & P(S=s|\boldsymbol{X}_i) =  \frac{\text{exp}\{\boldsymbol{U}^t_{is}\boldsymbol{\beta}_{0}\}}{1+ \sum_{k\in \mathcal{S}}\text{exp}\{\boldsymbol{U}^t_{ik}\boldsymbol{\beta}_{0}\}} \\
    & P(S=0|\boldsymbol{X}_i) = \frac{1}{1+ \sum_{k\in \mathcal{S}}\text{exp}\{\boldsymbol{U}^t_{ik}\boldsymbol{\beta}_{0}\}}. 
\end{split}
\end{equation*}
for $s \in \mathcal{S}$ with $\sum_{s\in \{0,\mathcal{S}\}}  P(S=s|\boldsymbol{X}_i) =1$. So, we have that 
\begin{equation}\label{Eq: L1.2_R2_2}
\begin{split}
    \sum_{s\in \mathcal{S}} \Big|I(S_i=s) - &  \frac{\text{exp}\{\boldsymbol{U}^t_{is}\boldsymbol{\beta}_0\}}{1+ \sum_{k\in \mathcal{S}}\text{exp}\{\boldsymbol{U}^t_{ik}\boldsymbol{\beta}_0\}}\Big| = \sum_{s\in \mathcal{S}} \left|I(S_i=s) - P(S=s|\boldsymbol{X}_i)\right| \\
    = & (1-P(S=S_i|\boldsymbol{X}_i)) + \sum_{s\in \mathcal{S}; s\ne S_i} P(S=s|\boldsymbol{X}_i) \\
    = & (1-P(S=S_i|\boldsymbol{X}_i)) + (1-P(S=S_i|\boldsymbol{X}_i) - P(S=0|\boldsymbol{X}_i)) \\
    \leq & 2
\end{split}
\end{equation}
By Equations \ref{Eq: L1.2_R2_1} and \ref{Eq: L1.2_R2_2}, we have that 
\begin{equation*}
\begin{split}
    & || \psi((S_i,\boldsymbol{V}_i),\boldsymbol{\beta}_0)|| \leq   2||\boldsymbol{V}_{i}|| \\
    & \Rightarrow || \psi((S_i,\boldsymbol{V}_i),\boldsymbol{\beta}_0)||^2 \leq   4||\boldsymbol{V}_{i}||^2 \\
\end{split}
\end{equation*}
for all $i=1,2,...,n$. So, under Assumption (vii) we have 
\begin{equation*}
\begin{split}
    & \sum_{i=1}^n ||\psi((S_i,\boldsymbol{X}_i),\boldsymbol{\theta}_0)||^2 \leq    4\sum_{i=1}^n |||\boldsymbol{V}_{i}||^2 = O(nm)\\ 
    & \Rightarrow \sum_{i=1}^n E\left(||\psi((S_i,\boldsymbol{V}_i),\boldsymbol{\beta}_0)||^2\right) = O(nm)
\end{split}
\end{equation*}
and since $q$ is constant $O(nm) = O(nmq)$ and Condition R2 holds 

The forth condition is 
\begin{itemize}[label={}]
\item \textbf{R3}. There exists a sequence of $(mq)\times (mq)$ matrices, $\boldsymbol{D}_n$ with $$\text{lim inf}_{n \to \infty} \lambda_{min}(\boldsymbol{D}_n)> 0$$ such that for any $B >0$ and uniformly in $\boldsymbol{\alpha} \in H_{m} = \{\boldsymbol{\alpha} \in \mathbb{R}^{m} : ||\boldsymbol{\alpha}|| =1\}$
\begin{equation*}
    \sup_{||\boldsymbol{\beta}-\boldsymbol{\beta}_0|| \leq B\left(\frac{mq}{n}\right)^{\frac{1}{2}}}\left|\boldsymbol{\alpha}^t\sum_{i=1}^n E_{\boldsymbol{\beta}_0}\left[\psi((S_i,\boldsymbol{V}_i),\boldsymbol{\beta}) - \psi((S_i,\boldsymbol{V}_i), \boldsymbol{\beta}_0)\right] - n\boldsymbol{\alpha}^t \boldsymbol{D}_n(\boldsymbol{\beta}-\boldsymbol{\beta}_0)\right|
\end{equation*}
is $o ((nm)^{\frac{1}{2}})$. 
\end{itemize}

To show Condition R3 holds, note that 
\begin{equation*}
\begin{split}
   \psi((S_i,\boldsymbol{V}_i),\boldsymbol{\beta})  = \sum_{s\in \mathcal{S}}\psi_s((S_i,\boldsymbol{V}_i),\boldsymbol{\beta})
\end{split}
\end{equation*}
where each 
\begin{equation*}
\begin{split}
   \psi_s((S_i,\boldsymbol{V}_i),\boldsymbol{\theta}) = \left[I(S_i=s) - \frac{\text{exp}\{\boldsymbol{U}^t_{is}\boldsymbol{\beta}\}}{1+ \sum_{k\in \mathcal{S}}\text{exp}\{\boldsymbol{U}^t_{ik}\boldsymbol{\beta}\}}\right]\boldsymbol{U}_{is} \\
\end{split}
\end{equation*}
From Example 3 in He and Shao \cite{he2000parameters}, we have that
\begin{equation*}
\begin{split}
    \Bigg|\boldsymbol{\alpha}^t\sum_{i=1}^n & E_{\boldsymbol{\beta}_0}\left[\psi_s((S_i,\boldsymbol{V}_i),\boldsymbol{\beta}) - \psi_s((S_i,\boldsymbol{V}_i), \boldsymbol{\beta}_0)\right] - n\boldsymbol{\alpha}^t \boldsymbol{G}_{n,s}(\boldsymbol{\beta}-\boldsymbol{\beta}_0)\Bigg| \\
    & \leq (\boldsymbol{\beta}-\boldsymbol{\beta}_0)^t\sum_{i=1}^n |\boldsymbol{\alpha}^t\boldsymbol{U}_{is}|\boldsymbol{U}_{is}\boldsymbol{U}_{is}^t(\boldsymbol{\beta}-\boldsymbol{\beta}_0)
\end{split}
\end{equation*}
where 
\begin{equation*}
\boldsymbol{G}_{n,s} = \frac{1}{n} \sum_{i=1}^n \frac{\text{exp}\{\boldsymbol{U}^t_{is}\boldsymbol{\beta}\}}{1+ \sum_{k\in \mathcal{S}}\text{exp}\{\boldsymbol{U}^t_{ik}\}}\boldsymbol{U}_{i,s}\boldsymbol{\boldsymbol{U}}_{i,s}^t. 
\end{equation*}
Because $\boldsymbol{G}_n$ as defined in Condition (xi) is $\sum_{s\in \mathcal{S}} \boldsymbol{G}_{n,s}$, we have that 
\begin{equation*}
\begin{split}
    \Bigg|\boldsymbol{\alpha}^t\sum_{i=1}^n & E_{\boldsymbol{\beta}_0}\left[\psi((S_i,\boldsymbol{V}_i),\boldsymbol{\beta}) - \psi((S_i,\boldsymbol{V}_i), \boldsymbol{\beta}_0)\right] - n\boldsymbol{\alpha}^t \boldsymbol{G}_n(\boldsymbol{\beta}-\boldsymbol{\beta}_0)\Bigg| \\
     \leq &  \sum_{s\in \mathcal{S}} \left|\boldsymbol{\alpha}^t\sum_{i=1}^n E_{\boldsymbol{\beta}_0}\left[\psi_s((S_i,\boldsymbol{V}_i),\boldsymbol{\beta}) - \psi_s((S_i,\boldsymbol{V}_i), \boldsymbol{\beta}_0)\right] - n\boldsymbol{\alpha}^t \boldsymbol{G}_{n,s}(\boldsymbol{\beta}-\boldsymbol{\beta}_0)\right| \\
     \leq & (\boldsymbol{\beta}-\boldsymbol{\beta})^t\sum_{i=1}^n \sum_{s\in \mathcal{S}} |\boldsymbol{\alpha}^t\boldsymbol{U}_{is}|\boldsymbol{U}_{is}\boldsymbol{U}_{is}^t(\boldsymbol{\beta}-\boldsymbol{\beta}) \\
     = & \sum_{i=1}^n \sum_{s\in \mathcal{S}} |\boldsymbol{\alpha}^t\boldsymbol{U}_{is}|\left[\boldsymbol{U}_{is}^t(\boldsymbol{\beta}-\boldsymbol{\beta})\right]^2 \\
\end{split}
\end{equation*}
By Condition (ix) we have that 
\begin{equation*}
\begin{split}
    \sup_{\boldsymbol{\alpha},\boldsymbol{\gamma} \in H_{qm}} \sum_{i=1}^n\sum_{s\in \mathcal{S}} (\boldsymbol{\alpha}^t \boldsymbol{U}_{is})^2 (\boldsymbol{\gamma}^t \boldsymbol{U}_{is})^2 \leq \sum_{s\in \mathcal{S}} \left[\sup_{\boldsymbol{\alpha},\boldsymbol{\gamma} \in H_{pm}} \sum_{i=1}^n (\boldsymbol{\alpha}^t \boldsymbol{U}_{is})^2 (\boldsymbol{\gamma}^t \boldsymbol{U}_{is})^2 \right] =O(nm)
\end{split}
\end{equation*}
Then, by results from Example 1 in \cite{he2000parameters}, we have that
\begin{equation*}
    \sup_{||\boldsymbol{\beta}-\boldsymbol{\beta}_0|| \leq B\left(\frac{mp}{n}\right)^{\frac{1}{2}}}\left|\boldsymbol{\alpha}^t\sum_{i=1}^n E_{\boldsymbol{\beta}_0}\left[\psi((S_i,\boldsymbol{V}_i),\boldsymbol{\beta}) - \psi((S_i,\boldsymbol{V}_i), \boldsymbol{\beta}_0)\right] - n\boldsymbol{\alpha}^t \boldsymbol{G}_n(\boldsymbol{\beta}-\boldsymbol{\beta}_0)\right| 
\end{equation*}
is $o ((nm)^{\frac{1}{2}})$ and Condition R3 holds. 

The fifth and sixth conditions are  
\begin{itemize}[label={}]
\item \textbf{R4}. For any $\boldsymbol{\beta} \in \mathbb{R}^{mq}$, $\boldsymbol{\alpha} \in H_{qm}$, and $B >0$,
\begin{equation*}
    \sup_{\boldsymbol{\tau}: ||\boldsymbol{\tau} - \boldsymbol{\theta}|| \leq B\left(\frac{qm}{n}\right)^{\frac{1}{2}}} \left\{\sum_{i=1}^n E_{\boldsymbol{\theta}}\left[|\boldsymbol{\alpha}^t\eta_i(\boldsymbol{\tau},\boldsymbol{\beta})|^2\right]\right\} = O(m)
\end{equation*}
\item \textbf{R5}. For any $\boldsymbol{\beta} \in \mathbb{R}^{mq}$ and $B >0$,
\begin{equation*}
    \sup_{\boldsymbol{\alpha} \in H_{qm}}\left\{\sup_{\boldsymbol{\tau}: ||\boldsymbol{\tau} - \boldsymbol{\beta}|| \leq B\left(\frac{qm}{n}\right)^{\frac{1}{2}}}\sum_{i=1}^n\left[\boldsymbol{\alpha}^t \eta_i(\boldsymbol{\tau},\boldsymbol{\theta}) \right]^2\right\} = O_P(m)
\end{equation*}
\end{itemize}
and follow from arguments very similar to those used in the proof for for Sub-Lemma 2.1 based on the bound on $||\eta_i(\boldsymbol{\tau}, \boldsymbol{\beta})||$ established in Equation \ref{Eq: MN_Bount_Eta}.

\qed

\newpage

\section{Appendix C: Additional Simulation Details} \label{sims.app}

As a reminder, the general simulation settings we considered are:

\begin{table}[!h]
    \centering
    \footnotesize
    \begin{tabular}{c|ccc}
    \hline
       \multirow{2}{*}{Setting} & \multirow{2}{*}{Study Sizes} & Outcome  & Main Source of Treatment \\
        & & Coefficients & Effect Heterogeneity \\
        \hline
        \hline
        1 & Similar & Random & Measured Features \\
        \hline
        2 & Different & Random & Measured Features \\
        \hline
        3 & Different & Random & Unmeasured Features \\
        \hline
    \end{tabular}
\end{table}

\subsection{Approximate Study Sizes}

When studies were of similar sizes, we used a numeric algorithm to solve for study-specific intercepts that resulted in, on average, a target population sample of around 5,500 participants and studies of around 1,500 participants \cite{robertson2022using}. In settings where we had three studies of different sizes, the target population sample was again around 5,500 participants and the studies, from Study 1 to Study 3, were around 675, 1350, and 2475 participants. In settings where we had thirty studies, we generated approximate study sizes by sampling thirty values from a $\text{Unif}(0,1)$ distribution, dividing each by the sum of the thirty values (so they sum to one), sorting them from smallest to largest, and assigning these to be the marginal probabilities of study membership conditional on not being in the target population sample for Studies 1 to 30. This resulting in approximate study sizes, from Study 1 to Study 30, of: 265, 271, 287, 377, 478, 557, 596, 693, 695, 766, 810, 829, 899, 987, 1450, 1472, 1730, 2023, 2122, 2160, 2161, 2194, 2433, 2433, 2436, 2575, 2619, 2830, 2865, and 2988.

\subsection{Coefficients for Study Membership Model}
In our simulations, observations were assigned to either enrollment in a study $s \in \mathcal{S} = \{1,2,...,m\}$ or membership in the target population using a multinomial logistic regression model where 
     \begin{equation*}
    \begin{split}
        & P(S_i=s|X_i) = \text{expit}\left(\beta_{s,0} +X_i \beta_{s,1}\right) \text{, for } s =1,2,...,m \\
        & P(S_i=0|X_i) = 1 - \sum_{s=1}^{m}P(S_i=s|X_i).
    \end{split}
    \end{equation*}

When $m=3$, the values of the $\beta_{s,1}$ we used, from Study 1 to Study 3, are: -0.4, -0.186, 0. Under Setting 1, we the values of the $\beta_{s,0}$ we used, from Study 1 to Study 3, are: -1.347, -1.302, -1.299. Under settings 2 and 3, we instead used: -2.16, -1.414, -0.799. 

When $m=30$, the values of the $\beta_{s,1}$ we used, from Study 1 to Study 30, are: -0.35, -0.34, -0.338, -0.334, -0.329, -0.293, -0.29, -0.289, -0.284, -0.281, -0.276, -0.244, -0.238, -0.222, -0.206, -0.167, -0.145, -0.142, -0.141, -0.135, -0.132, -0.102, -0.081, -0.071, -0.067, -0.066, -0.045, -0.019, -0.016, -0.002. Under Setting 1, we the values of the $\beta_{s,0}$ we used, from Study 1 to Study 30, are: -1.304, -1.302, -1.302, -1.301, -1.3, -1.295, -1.294, -1.294, -1.294, -1.293, -1.293, -1.29, -1.289, -1.288, -1.287, -1.286, -1.286, -1.286, -1.286, -1.287, -1.287, -1.288, -1.289, -1.29, -1.291, -1.291, -1.293, -1.296, -1.297, -1.299. Under settings 2 and 3, we instead used: -3.058, -3.033, -2.976, -2.7, -2.463, -2.303, -2.235, -2.084, -2.08, -1.982, -1.925, -1.897, -1.815, -1.719, -1.333, -1.315, -1.152, -0.995, -0.948, -0.93, -0.929, -0.913, -0.811, -0.811, -0.81, -0.754, -0.738, -0.663, -0.65, -0.61. 

\subsection{TATEs Used to Evaluate Performance} \label{Append: exptarget_details}

In order to evaluate the performance of the estimators, we had to be able to calculate the true TATE in the target population. This requires knowing $E(X|S=0)$. To obtain a close approximation to this value, we conducted simulations with 50,000 replications. The resulting approximations to the true TATEs, as well as the approximations to $E(X|S=0)$ used to obtain them, in each setting are:

\begin{table}[!h]
\centering
\begin{tabular}{c|c|ccc}
  \hline
 $m$ & Settings & TATE & $E(X|S=0)$ & MCSE of $E(X|S=0)$ \\ 
  \hline
  \hline
   \multirow{2}{*}{3} & 1 & -1.04 & 0.09 & $5.9 \times 10^{-5}$ \\ 
    & 2 and 3 & -1.03 & 0.05 & $5.9 \times 10^{-5}$ \\ 
     \hline
   \multirow{2}{*}{30} & 1 & -1.08 & 0.17 & $6 \times 10^{-5}$ \\ 
   & 2 and 3 & -1.05 & 0.11 & $6 \times 10^{-5}$ \\ 
   \hline
\end{tabular}
\end{table}

\end{document}